\documentclass[fleqn,12pt,twoside]{article}
\usepackage[headings]{espcrc1}

\readRCS
$Id: espcrc1.tex,v 1.2 2004/02/24 11:22:11 spepping Exp $
\usepackage{graphicx}
\usepackage[figuresright]{rotating}

\newcommand{\AmS}{{\protect\the\textfont2
  A\kern-.1667em\lower.5ex\hbox{M}\kern-.125emS}}

\newcommand{\AuAu}{\mbox{Au+Au}}
\newcommand{\CuCu}{\mbox{Cu+Cu}}

\newcommand{\pizero}{\mbox{$\pi^0$}}
\newcommand{\Jpsi}{\mbox{$J/\psi$}}
\newcommand{\ee}{\mbox{$e^+e^-$}}
\newcommand{\mumu}{\mbox{$\mu^+\mu^-$}}
\newcommand{\pt}{\mbox{$p_T$}}
\newcommand{\pttrig}{\mbox{$p_T^{\rm trig}$}}
\newcommand{\Ettrig}{\mbox{$E_T^{\rm trig}$}}
\newcommand{\ptassoc}{\mbox{$p_T^{\rm assoc}$}}
\newcommand{\Npart}{\mbox{$N_{\rm part}$}}

\newcommand{\RAA}{\mbox{$R_{\rm AA}$}}
\newcommand{\RCP}{\mbox{$R_{\rm CP}$}}
\newcommand{\vtwo}{\mbox{$v_2$}}

\newcommand{\rootsnn}{\mbox{$\sqrt{s_{NN}}$}}
\newcommand{\etal}{\mbox{\it et al.}}

\title{Quark Matter 2005: Experimental Conference Summary}

\author{Itzhak Tserruya
            \address[MCSD]{Department of Particle Physics, Weizmann Institute of Science, Rehovot 76100, Israel}
         }

\runtitle{Experimental Conference Summary}
\runauthor{Itzhak Tserruya}

\begin{document}

\maketitle

\begin{abstract}
    Highlights of the experimental results presented at the Quark Matter 2005 Conference in Budapest (Hungary) are reviewed
    and open issues are discussed.
\end{abstract}

\section{Introduction}
\label{sec:intro}
  This was an excellent conference. For the fourth consecutive time since the beginning of RHIC operation in the year 2000, the
Quark Matter conference displayed the strong vitality and enormous productivity of the field. A few facts and numbers easily 
prove this point. Since the last Conference in Oakland 19 months ago, (i) there have been two very successful additional runs at 
RHIC, run-4 and run-5, basically completing a broad survey of the physics landscape not only in Au+Au collisions at the top RHIC 
energy of $\sqrt{s_{NN}}$=200 GeV but also at a lower energy, 62 GeV and in a lighter system, Cu+Cu at $\sqrt{s_{NN}}$=200, 62 
and 22.5 GeV, and including the crucial reference runs p+p and d+Au at 200 GeV, (ii) I counted 87 papers published in the 
refereed literature, 30 of them in Phys. Rev. Lett., (iii) monumental White Papers \cite{white-papers} based on the first three 
years of RHIC operation, were published by the four RHIC experiments reflecting their assessment on the nature of the matter 
formed at RHIC and a new logo was born, the "perfect fluid". All this and numerous new experimental results from RHIC and SPS 
formed the core of the 25 and 60 experimental talks given in plenary and parallel sessions, respectively, which I shall try to 
summarize here.
 
The last three Quark Matter conferences were naturally dominated by RHIC data, more precisely RHIC data on global event
characterization (multiplicity, flow, HBT, fluctuations, kinetic and chemical equilibrium) and the newly discovered high p$_T$
phenomena. This conference was also dominated by RHIC data. But with the advent of results from the high luminosity runs, the 
emphasis moved to penetrating and rare probes (charm, charmonium, dileptons and photons). High p$_T$ phenomena and jet physics 
remained a very strong topic. It is also noteworthy that first results were presented from the Cu+Cu run-5 which ended just a 
few months prior of the conference. The timely release of results is an impressive feature characteristic of all RHIC runs so 
far. This remarkable achievement would not be possible without the dedication and enthusiasm of many young (and not so young) 
and brilliant researchers who for several months worked literally around the clock, forgetting week-ends and other obligations, 
to bring their analyses to the high level required at this conference. I wish to pay tribute to all of them here.

In addition to the RHIC data, we have also seen results from NA60, a second generation experiment at the CERN SPS devoted to the
measurement of the dimuon spectrum from the low-masses up to the J/$\psi$. With its two main characteristic features, excellent
mass resolution and high statistics, NA60 presented superb data from its first run (that unfortunately also seems to be its last
one) on In+In at 158~AGeV.

In coming to summarize this conference, it is obviously impossible to do justice to all the new results presented. This
summary presents a broad choice, but it is a personal one, and I apologize to all the speakers whose results I could
not include in the limited space of this summary.

\section{Elliptic Flow}
\label{sec:flow}
\begin{figure}[htb]
\begin{minipage}{0.65\textwidth}
     \vspace*{-0.7cm}
     \includegraphics*[width=9cm, height=6cm]{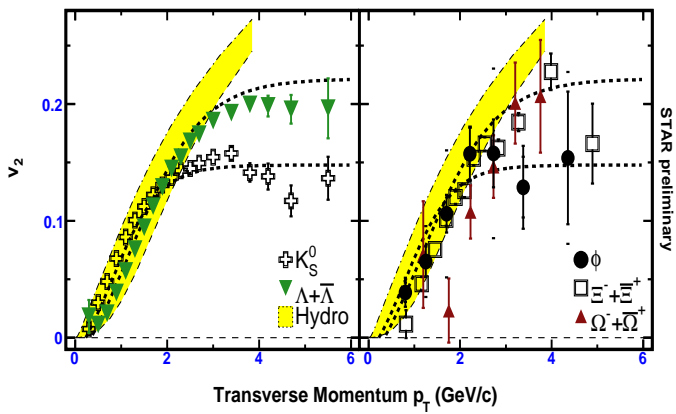}
\end{minipage}
\begin{minipage}{0.35\textwidth}
     \vspace*{-2.0cm}
     \caption{Elliptic flow parameter $v_{2}$ for strange (left) and multi-strange
              (right) hadrons measured by STAR in Au+Au collisions at
              \rootsnn = 200 GeV. The curves are empirical fits. Hydrodynamic
              calculations are shown by the shaded areas \cite{star-v2oldenburg}.}
\label{fig:star-wangfig4}
\end{minipage}
\vspace*{-0.5cm}
\end{figure}

  We have seen at this conference many new results on flow mainly from RHIC but also from SPS.
Perhaps the most striking result is that open charm, inferred from measurements of inclusive electrons after subtraction of 
hadron decays, exhibits elliptic flow of comparable magnitude to that of hadrons. This result and the challenge to energy loss 
models will be discussed in Section~\ref{sec:heavy-flavor}.
Elliptic flow measurements at RHIC have been reported for many new particles, including heavy and rare particles, in particular 
d and multistrange hadrons, $\phi$, $\Xi$ and $\Omega$, (see Fig.~\ref{fig:star-wangfig4}) \cite{star-wang,phenix-vicki}. So 
far, {\it all} hadrons measured at RHIC show elliptic flow. The only notable particle where no results are as yet available is 
the \Jpsi. I shall return to the importance of this missing measurement again in Section~\ref{sec:Jpsi}.

\begin{figure}[ht]
\begin{minipage}{0.65\textwidth}
     \vspace*{-0.7cm}
     \includegraphics*[width=9cm, height=6.5cm]{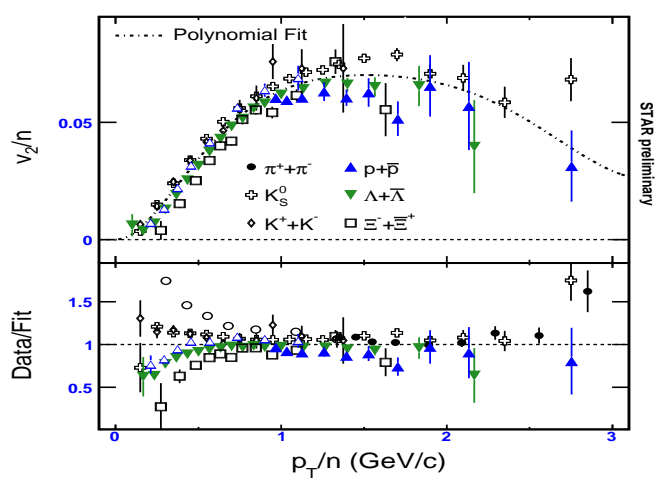}
\end{minipage}
\begin{minipage}{0.35\textwidth}
     \vspace*{-2.0cm}
     \caption{\vtwo $/n$ vs. \pt $/n$ where $n$ is the number of valence quarks,
              illustrating the constituent quark scaling of elliptic
              flow. The bottom panel shows the deviations of the measurements
              from an empirical fit through all data points \cite{star-wang}.}
\end{minipage}
\label{fig:star-wangfig6} \vspace*{-0.8cm}
\end{figure}

All the new RHIC results follow and confirm the pattern already established over the last few years from the wealth of data 
available on this topic (see the comprehensive review of Lacey at this conference \cite{lacey-flow-review}): at low \pt\ (\pt\ 
smaller than $\sim$ 2 GeV/c) \vtwo\ increases monotonically with \pt\ and the results are well reproduced by hydrodynamic 
calculations. In the "intermediate" \pt\ region (\pt\ = 2-5 GeV/c) the elliptic flow flattens and the hydrodynamic description 
breaks down as shown e.g. in Fig.~\ref{fig:star-wangfig4}. The data in this range follow remarkably well the scaling with the 
number of valence quarks predicted by recombination models \cite{duke,molnar-voloshin} \footnote{quark scaling is predicted in 
the limiting case where the combining quarks carry the same fraction of the hadron momentum.}. This is illustrated in 
Fig.~\ref{fig:star-wangfig6} which shows a comprehensive compilation of elliptic flow measurements  with such a scaling. 
Deviations are seen at low \pt\ where the hydrodynamic description works best but at intermediate \pt\ quark scaling works. This 
suggests that quarks are the relevant degrees of freedom very early in the collision, when flow takes place. However, it should 
be stressed that these are constituent quarks.

The measurements of \vtwo\ have been extended to higher \pt\ values \cite{phenix-winter,star-v2}. \vtwo\ decreases but retains a 
non-zero value up to \pt\ = 10 GeV/c, a behavior that is consistent with energy loss models.

The new RHIC flow results contribute to strengthening the case for a strongly interacting quark-gluon plasma, sQGP, 
characterized by early thermalization of partonic matter made of constituent quarks, behaving like a perfect fluid. 

SPS is catching up on flow. $\Lambda$ \vtwo\ results, from measurements done in the year 2000 at 158 AGeV, were shown by NA49 
\cite{na49-stefanek-flow} and CERES \cite{ceres-milosevic-flow}. The two are in good agreement with each other. At low \pt, 
\vtwo\ increases monotonically, as at RHIC energies, but the absolute values are lower at the SPS and hydrodynamic calculations 
cannot simultaneously reproduce particle spectra and flow. When constrained to reproduce the particle spectra, the hydrodynamic 
calculations overestimate the amount of flow.

\section{Varying the system size: Cu+Cu vs. Au+Au}
\label{sec:system-size}
  What new information could be gained by studying different systems at the same energy? This question has come up repeatedly over
the years. Proponents have proposed this as an additional knob to turn for the systematic study of relativistic heavy-ion 
collisions. Clearly the initial geometrical overlap volume in central Cu+Cu collisions has not the same shape as in Au+Au 
collisions at the selected centrality corresponding to the same number of participants. But if the system quickly expands and 
reaches thermal equilibrium, this difference could rapidly be erased from the collision memory. Therefore opponents have argued 
that going to lighter systems would be equivalent to study a heavier one as function of centrality.

\begin{figure}[h]
   \vspace*{-0.8cm}
   \begin{minipage}[t]{70mm}
      \includegraphics*[width=7.5cm, height=6.7cm]{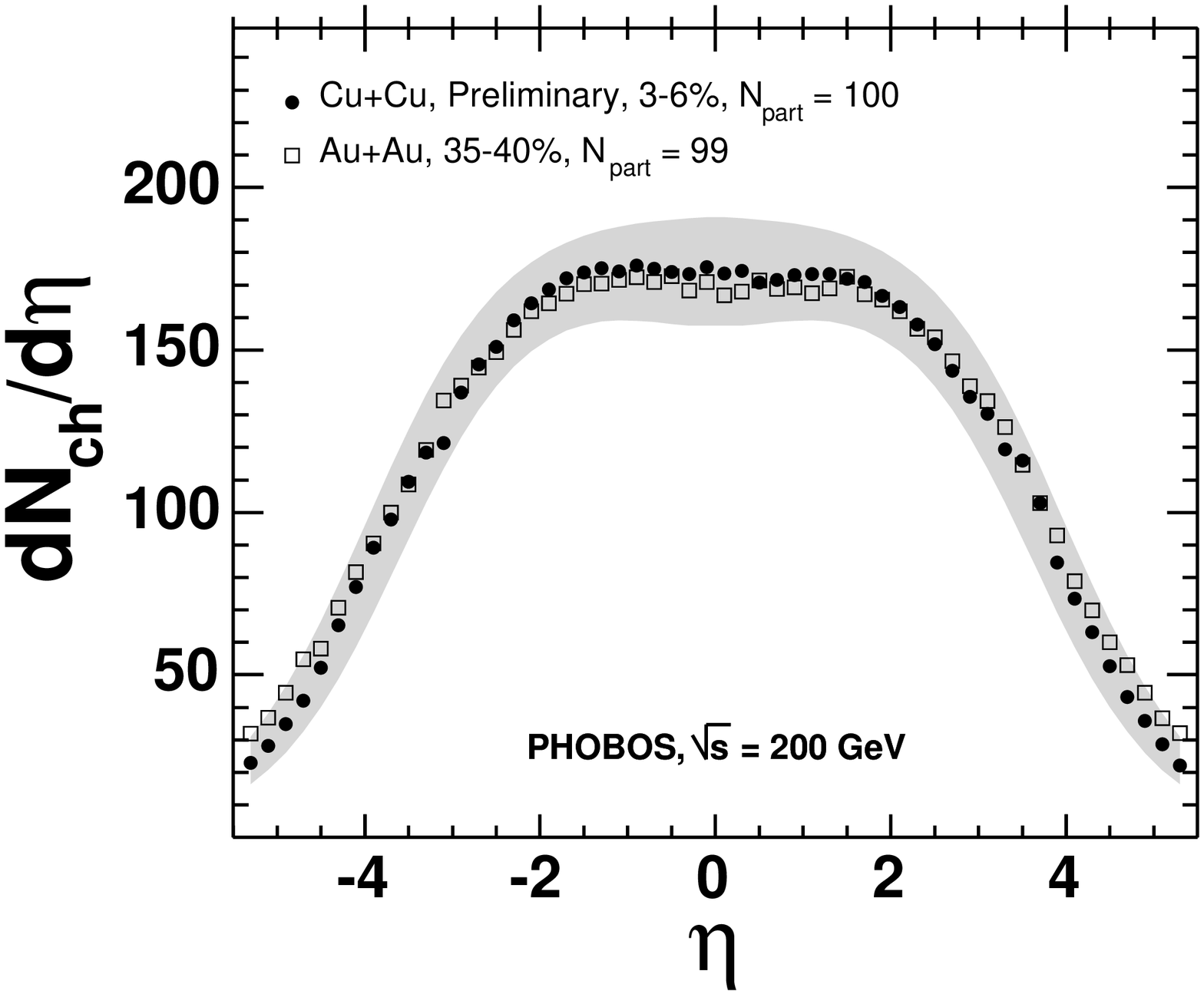}
      \vspace*{-1.7cm}
      \caption{Charged hadron pseudorapidity distribution in \rootsnn\ = 200 GeV \CuCu\
               and \AuAu\ collisions with similar \Npart\ \cite{phobos-roland}.}
      \label{fig:phobos-roland5}
   \end{minipage}
      \hspace{\fill}
   \begin{minipage}[t]{70mm}
      \includegraphics[width=7.5cm]{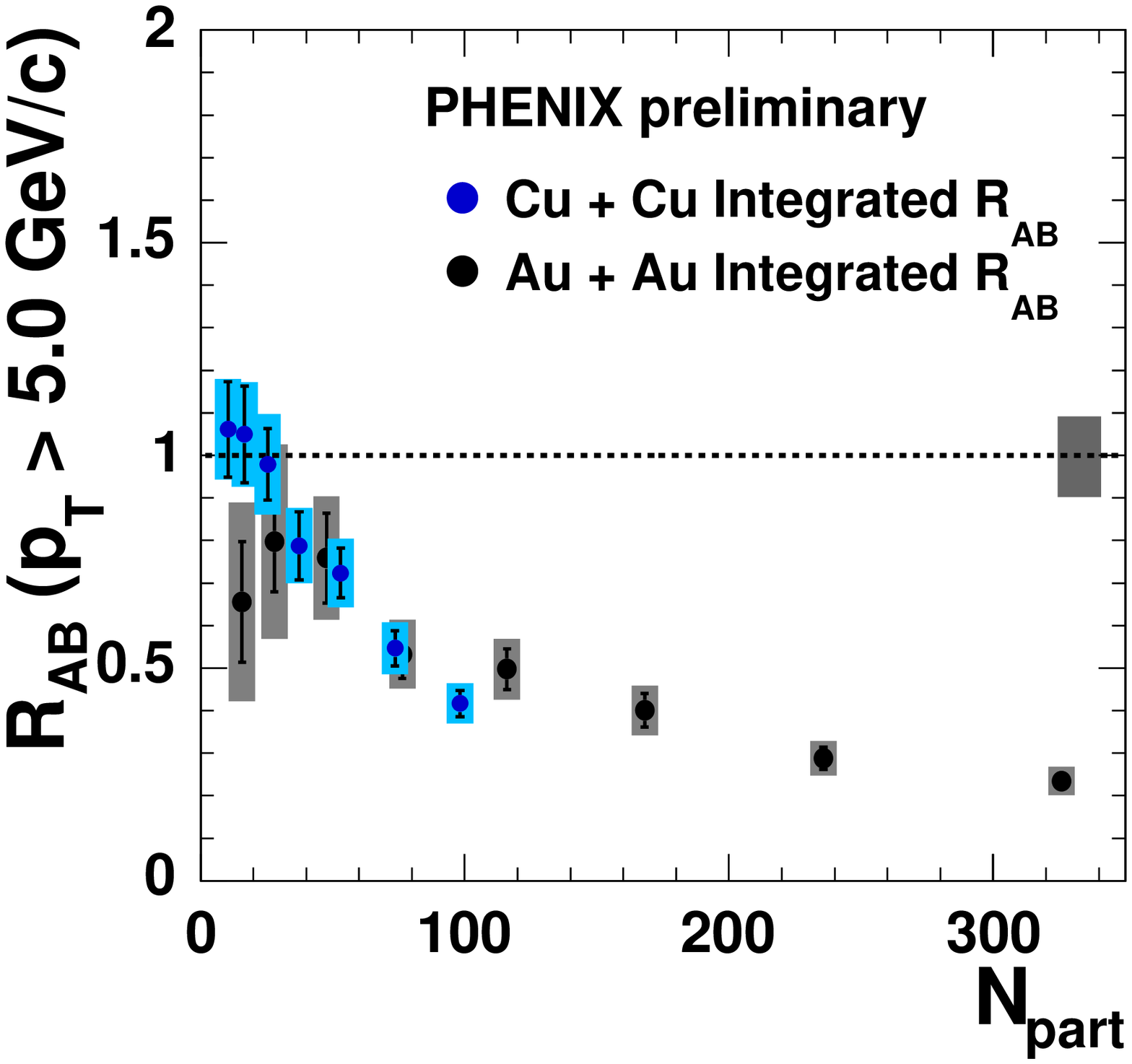}
      \vspace*{-1.7cm}
      \caption{Nuclear modification factor \RAA\ for \pizero\ vs.
               \Npart\ in central Cu+Cu and Au+Au collisions at \rootsnn\ = 200 GeV \cite{phenix-vicki}.}
      \label{fig:phenix-vicki8}
   \end{minipage} \vspace*{-0.5cm}
\end{figure}

  The many new results from the recent Cu+Cu run at RHIC  provided a clear answer: Cu+Cu is very similar to Au+Au
when the two systems are compared for collisions with the same number of participants. A very nice example is shown in
Fig.~\ref{fig:phobos-roland5} which compares the charged hadron pseudorapidity distributions measured by PHOBOS in Cu+Cu and 
Au+Au collisions at \rootsnn\ =200 GeV \cite{phobos-roland}. The two distributions are almost identical when the centrality bins 
are chosen to correspond to the same number of participants. The same holds true for other centralities and also at \rootsnn\ = 
62.4 GeV \cite{phobos-roland}.

Many other examples supporting this argument were provided by the four RHIC experiments (see
\cite{phenix-vicki,phobos-roland,star-dunlop,brahms-staszel}). I shall mention two more: PHOBOS \cite{phobos-roland}
showed that the elliptic flow \vtwo\ relative to the participant eccentricity scales with \Npart\ irrespective of the
collision system (see also \cite{phenix-masui-flow}). The \Jpsi\ production also scales with the number of participants
and this was demonstrated by the PHENIX results obtained in Au+Au and Cu+Cu collisions (see Section~\ref{sec:Jpsi}) and
also at SPS in a comparison of the Pb+Pb and In+In results from NA50 and NA60, respectively.

There is however a real intrinsic benefit in the study of a light system. It adds significant precision to the determination of 
\Npart\ for \Npart $\leq$ 100. This is illustrated in Fig.~\ref{fig:phenix-vicki8} which compares the nuclear modification 
factor \RAA\ of \pizero\ in   \CuCu\ and \AuAu\ collisions vs. \Npart \cite{phenix-vicki}. The same behavior is observed in the 
two systems for the same \Npart\ but the systematic uncertainties are significantly lower in \CuCu\ (see also 
\cite{phobos-roland,star-dunlop}).

\section{High p$_T$ hadrons}
\label{sec:high-pt}
   The suppression of high \pt\ hadrons in central Au+Au collisions is among the unique and most significant
phenomena discovered at RHIC. With the high luminosities of run-4 and run-5 it became possible to extend the \pt\ reach to much
higher values and to perform more detailed and higher precision studies. The new data presented at the conference reinforce the
current interpretation in terms of energy loss in a dense and opaque medium but also raise questions and problems about the
energy loss mechanisms (see the dedicated reviews on high \pt\ physics \cite{jacobs,cole}) in particular the very intriguing and
challenging results on charm suppression which will be discussed in Section~\ref{sec:heavy-flavor}.

\begin{figure}[h]
   \begin{minipage}[t]{70mm}
      \includegraphics*[width=7.5cm,height=5.4cm]{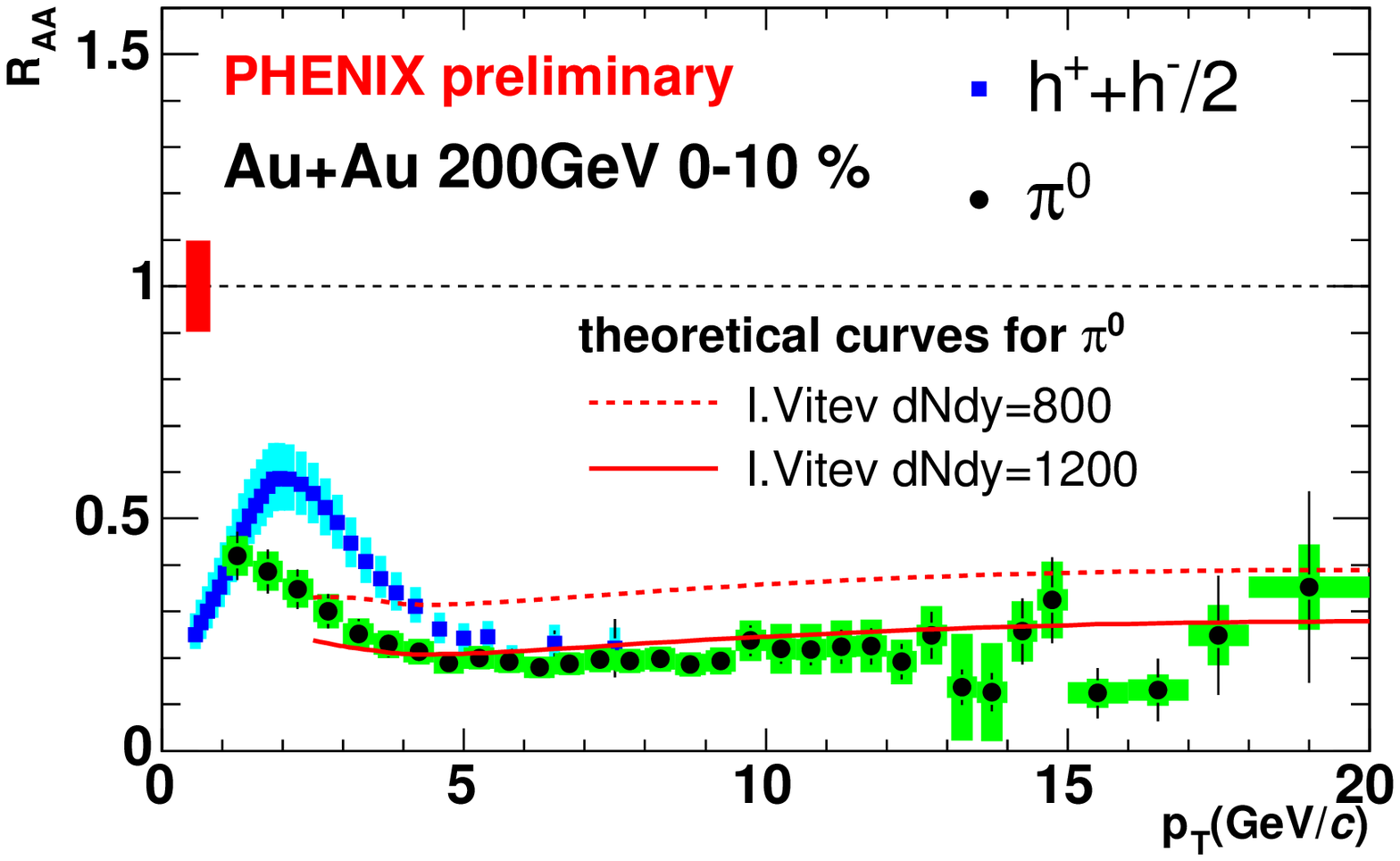}
      \vspace*{-1.7cm}
      \caption{\RAA\ for \pizero\ and charged hadrons in central Au+Au collisions from PHENIX \cite{phenix-shimomura}
               together with radiative energy loss calculations \cite{vitev}.}
    \label{fig:phenix-shimomura1}
   \end{minipage}
      \hspace{\fill}
   \begin{minipage}[t]{70mm}
      \includegraphics[width=7.5cm,height=5.4cm]{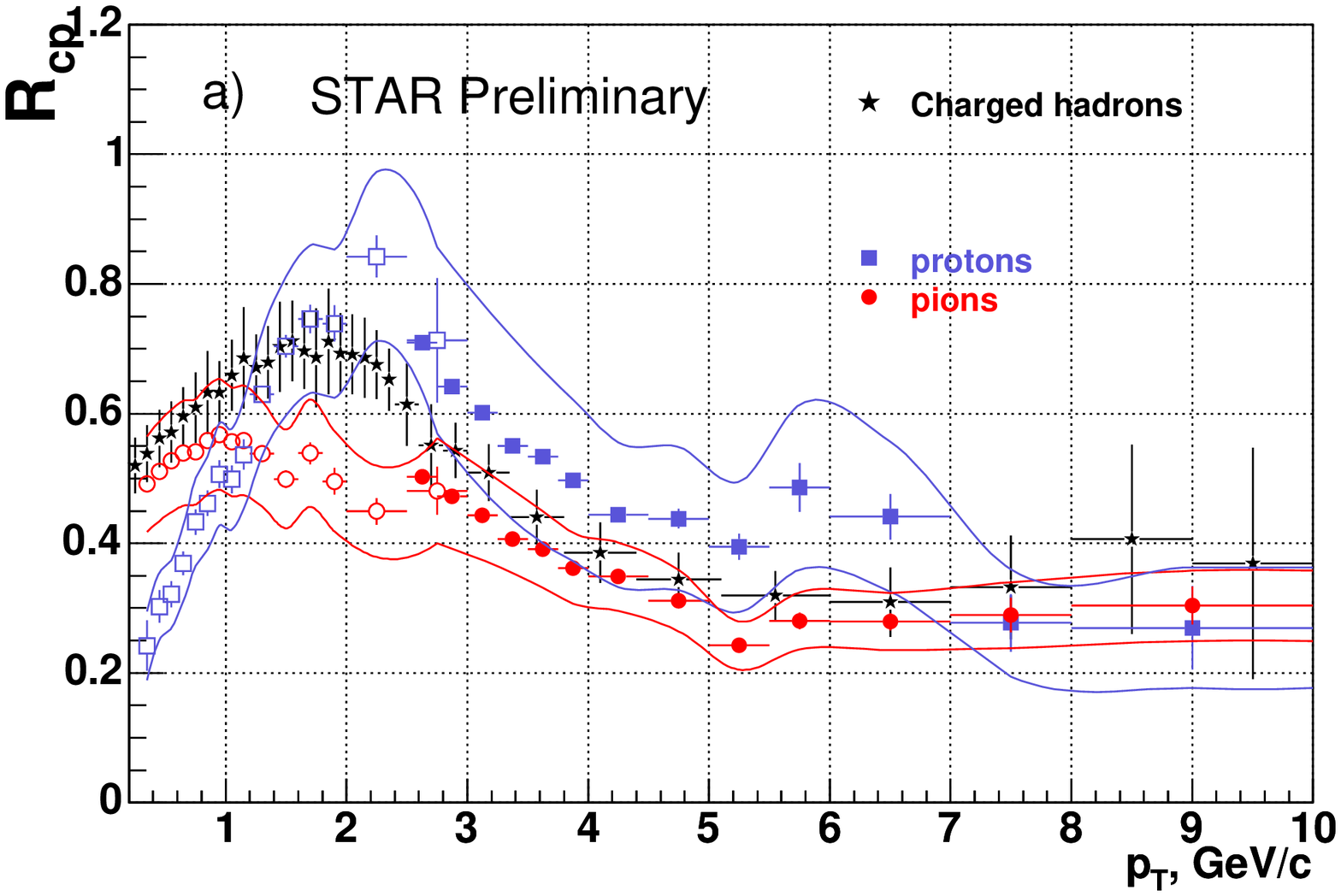}
      \vspace*{-1.7cm}
      \caption{\RCP, the ratio of $N_{binary}$-scaled \pt\ spectra between central (0-5\%) and peripheral (60-80\%) Au+Au
               collisions at \rootsnn\ = 200 GeV of $\pi,p$ and charged hadrons \cite{star-barannikova}.}
      \label{fig:star-barannikova1}
   \end{minipage}
    \vspace*{-0.8cm}
\end{figure}

Fig.~\ref{fig:phenix-shimomura1} shows the nuclear modification factor of \pizero\ measured by PHENIX in central Au+Au 
collisions at \rootsnn\ = 200 GeV \cite{phenix-shimomura}. The suppression is very strong, with \RAA\ $\sim$0.2, and remains 
flat up to 20~GeV/c. Very similar results were shown for Cu+Cu \cite{phenix-shimomura}. This behavior is well reproduced by 
partonic radiative energy loss models \cite{vitev,wang04,eskola}. The figure shows one such calculation implying an average 
gluon density $dN_g/dy \sim$ 1200 \cite{vitev}. The figure also displays the \RAA\ of charged hadrons. The difference between 
charged hadrons and \pizero's in the "intermediate" \pt\ region (\pt\ $\sim$2-5 GeV/c), attributed to the proton contribution 
and explained by recombination models, disappears at $\pt
>$ 5~GeV/c in agreement with the same recombination models (not shown) \cite{duke,oregon,tamu}. Similar results are
shown in Fig.~\ref{fig:star-barannikova1} which compares the \RCP\ of $\pi$ and $p$ \cite{star-barannikova}.

\begin{figure}[ht]
  \vspace*{-0.5cm}
  \begin{minipage}[t]{70mm}
      \includegraphics*[width=7.5cm, height=6cm]{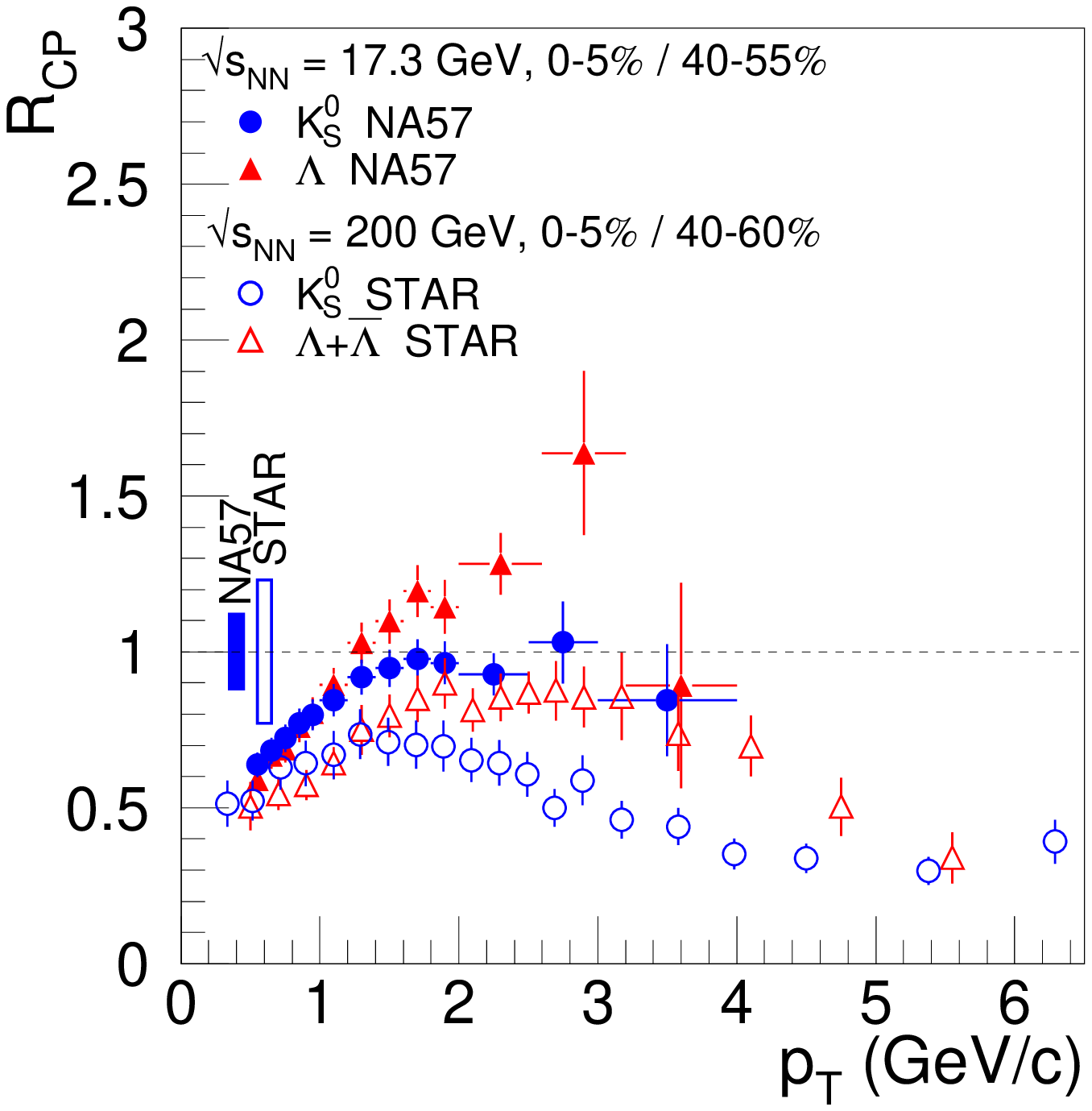}
  \end{minipage}
  \hspace{\fill}
  \begin{minipage}[t]{70mm}
     \includegraphics*[width=7cm, height=6cm]{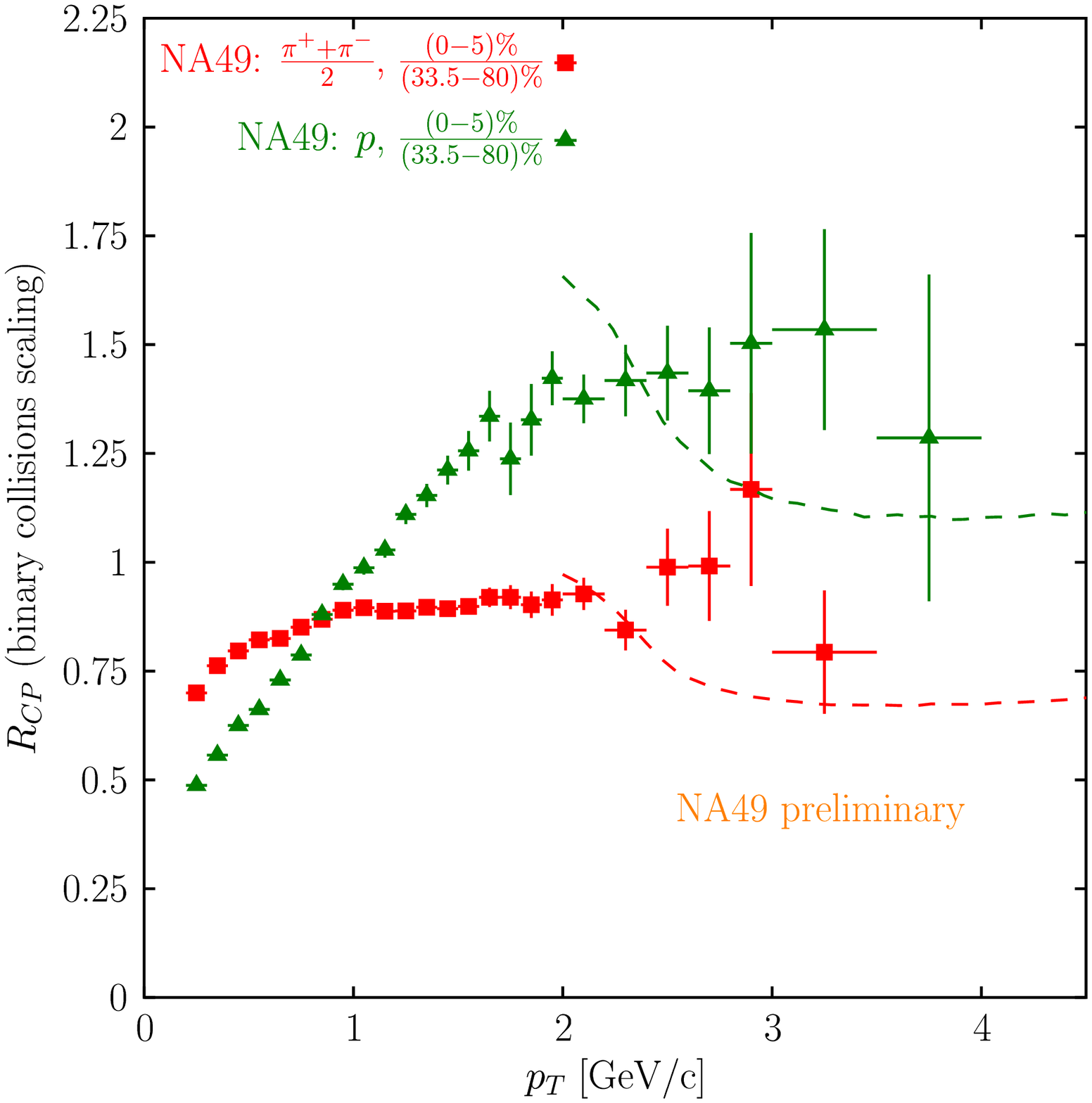}
  \end{minipage}
  \vspace*{-1.0cm}
  \caption{\RCP\ results from SPS. Left panel: results from NA57 for $K_S^0$ and $\Lambda$ compared to similar
            data from STAR \cite{na57-dainese}. Right panel: results from NA49 for $p$ and $\pi$ \cite{na49-laszlo}.}
   \label{fig:na57-NA49-rcp}
   \vspace*{-0.5cm}
\end{figure}

NA49 and NA57 presented \RCP\ results at the top SPS energy \rootsnn = 17.3 GeV providing much needed information on the energy 
dependence of high \pt\ suppression \cite{na49-laszlo,na57-dainese}. Results are shown in Fig.~\ref{fig:na57-NA49-rcp} for 
$K_S^0$ and $\Lambda$ from NA57 compared to similar data from STAR (left panel) and $\pi$ and $p$ from NA49 (right panel).
The same systematic behavior as in RHIC is observed, \RCP(baryon) $>$ \RCP(meson) and radiative energy loss models do quite well
in reproducing the SPS results. The high \pt\ suppression is much weaker than at RHIC but it is strong enough that the expected
Cronin enhancement at high \pt\ is not observed for mesons. The SPS data as well as the preliminary results obtained at the
intermediate RHIC energy of \rootsnn = 62.4 GeV \cite{star-salur} are reproduced with an initial gluon density of $dN_g/dy$ that
scales as the particle density  $dN_{ch}/dy$ \cite{wang04}. Like with many other observables, there seems to be a smooth energy
dependence of the high \pt\ hadron suppression from SPS to top RHIC energies.

All these results, from RHIC and SPS, appear as a great victory for the partonic energy loss models. However, new results 
presented at the conference will allow more stringent tests of the models. In particular, results on the dependence of the 
\pizero\ suppression on the emission angle with respect to the reaction plane, as function of \pt\ and centrality 
\cite{phenix-winter}, will be very valuable in elucidating the interplay of elliptic flow and energy loss as function of \pt\ 
and particularly in the intermediate \pt\ range where the energy loss models fail in reproducing the large values of \vtwo\ 
\cite{shuryak02}.

\begin{figure}[htb]
  \vspace*{-0.6cm}
  \centering
  \includegraphics*[width=14cm,height=5.5cm]{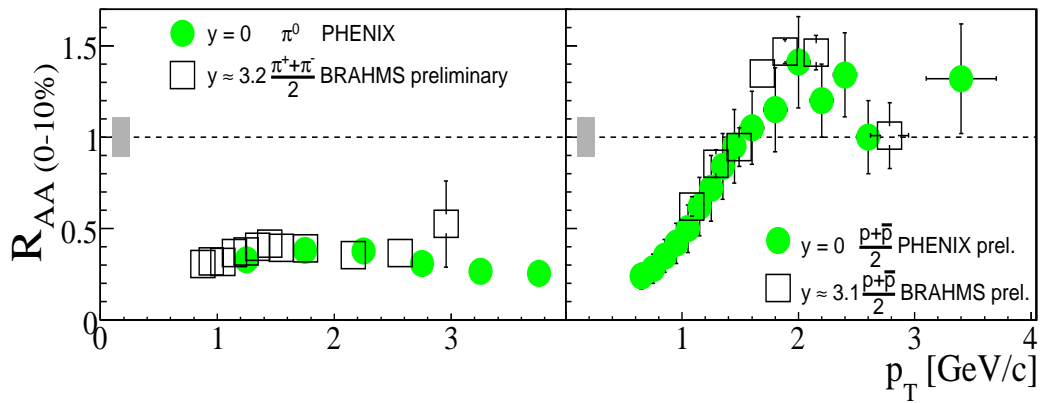}
  \vspace{-1.0cm}
  \caption{\RAA\ for $\pi$ (left panel) and $p$ (right panel) measured by BRAHMS at $y$ = 3.2 and by PHENIX at $y$ = 0 in
  central Au+Au collisions at \rootsnn = 200 GeV \cite{brahms-staszel}.}
  \label{fig:brahms-staszel13}
  \vspace*{-0.6cm}
\end{figure}

BRAHMS provided clear evidence that \RAA\ does not depend on rapidity as demonstrated in Fig.~\ref{fig:brahms-staszel13}. The 
figure shows excellent agreement between the \RAA\ for $\pi$ and p measured by BRAHMS at $y$ = 3.2 and same data measured by 
PHENIX at $y$ = 0 in central Au+Au collisions  \cite{brahms-staszel}. This is a surprising result and a challenge for theory. In 
the jet quenching models, the gluon density is smaller at forward rapidity and consequently one would expect a lower 
suppression. As a possible explanation one could argue that at forward rapidity, other mechanisms are present, like gluon 
saturation as manifest in d+Au measurements, that could lead to an additional suppression such that by accident the total 
suppression looks similar to the one at mid-rapidity \cite{xnw-private}. A quantitative analysis incorporating all effects in a 
comprehensive theoretical approach is needed.

\section{Jet correlations}
\label{sec:jets}
  With the availability of large data sets it is now possible to study particle correlations over a large range 
of \pt\ values both for the trigger as well as for the associated particles. Each selection of kinematic cuts provides a slice 
of the jet structure and reveals limited but valuable information. Combining all these selections together should eventually 
lead to a detailed reconstruction of the jet topology and its interaction with the medium allowing complete tomography of the 
matter created at RHIC.

First jet studies at RHIC provided the by now classical result of the away-side jet disappearance in central Au+Au at high \pt\ 
values ($\pttrig >4$ GeV/c and $\ptassoc > 2$ GeV/c) \cite{star-jet2003}. This dramatic result reflects the limited ability to 
extract the suppressed away-side jet from the high background in this particular kinematic cut rather than a real disappearance 
of the jet. If one includes softer associated particles (0.15 $< \ptassoc  < 2$ GeV/c while keeping the same trigger $\pttrig 
>4$ GeV/c), the correlation strength grows faster than the background and a clear and broad away-side peak appears as shown in 
recent studies \cite{star-jet2005}. From these two observations a picture emerges which is fully compatible with parton energy 
loss in a high-density medium: the hadron trigger preferentially selects a dijet produced near to the surface. The parton 
heading outwards gives rise to an almost unaffected near-side jet that fires the trigger whereas the recoiling parton looses 
energy while traversing the medium and its momentum gets redistributed over several particles, suppressing high \pt\ particles 
(disappearance of away side peak in \cite{star-jet2003}) and enhancing low \pt\ particles (broad away side peak in 
\cite{star-jet2005}).

\begin{figure}[h]
    \vspace*{-0.7cm}
    \centering
    \includegraphics[width=0.8\textwidth]{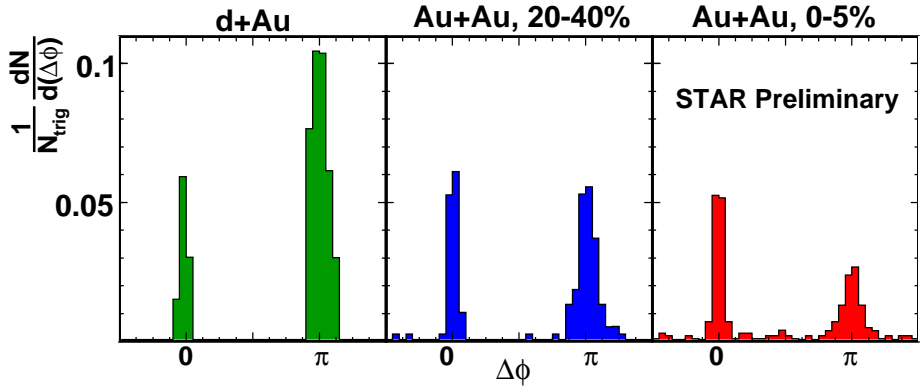}
     \vspace*{-0.9cm}
     \caption{Azimuthal correlations of charged hadrons (8 $< \pttrig < 15$ GeV) with associated particles 
             (6 GeV/c $ < \ptassoc < \pttrig$) in d+Au, semi-central (20-40\%) and central (0-5\%) Au+Au collisions \cite{star-magestro}.}
     \label{fig:star-magestro-c}
     \vspace*{-0.8cm}
\end{figure}

\begin{figure}[h]
   \vspace*{-0.6cm}
   \begin{minipage}[t]{70mm}
      \includegraphics[width=7.5cm,height=5.5cm]{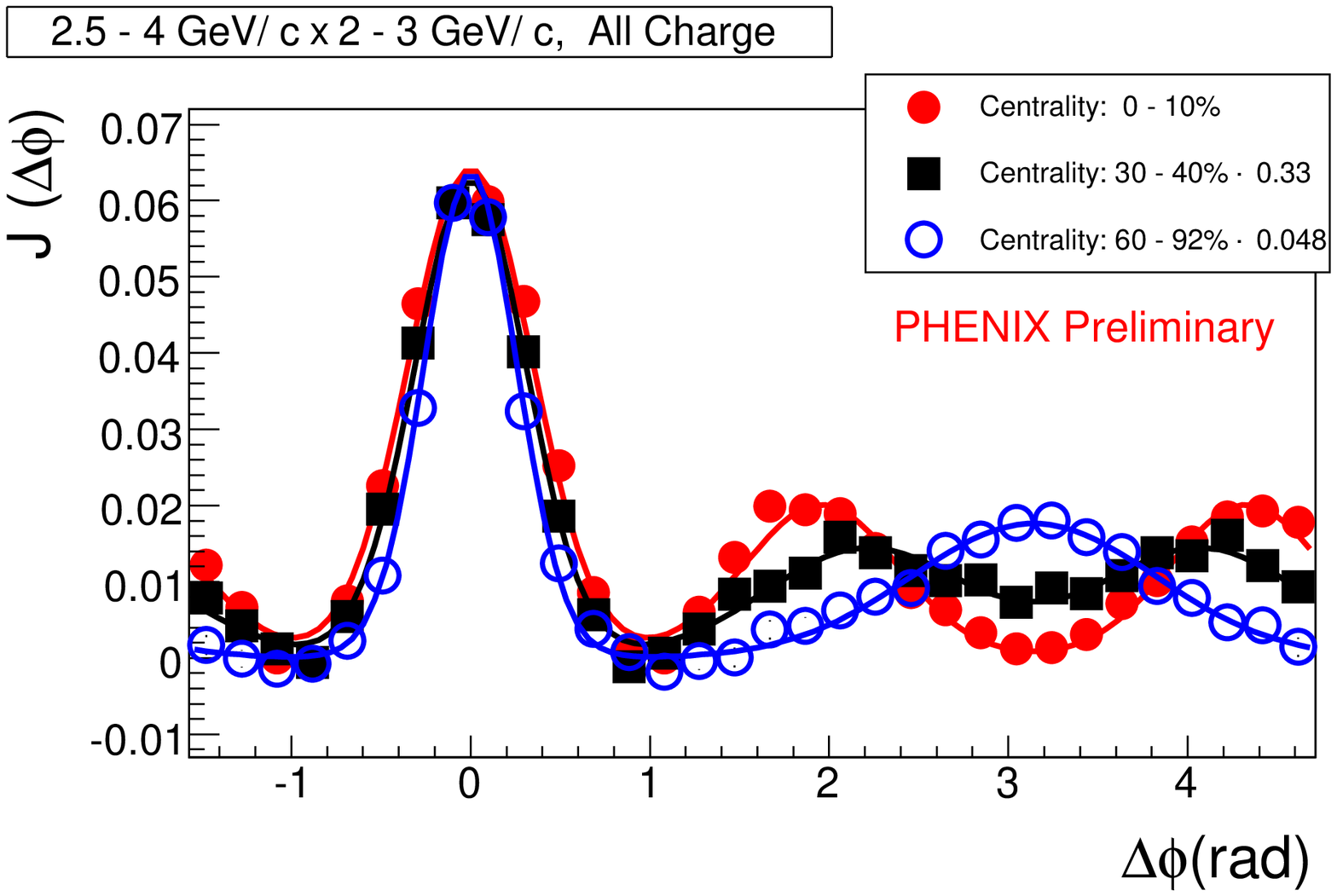}
      \vspace*{-1.7cm}
      \caption{Correlation function for central, mid-central, and peripheral Au+Au collisions at $\sqrt{s_{NN}} = 200$ GeV
            after subtraction of the flow contribution \cite{phenix-buesching}.}
   \label{fig:phenix-buesching4}
   \end{minipage}
      \hspace{\fill}
   \begin{minipage}[t]{70mm}
      \includegraphics[width=7.5cm,height=5.5cm]{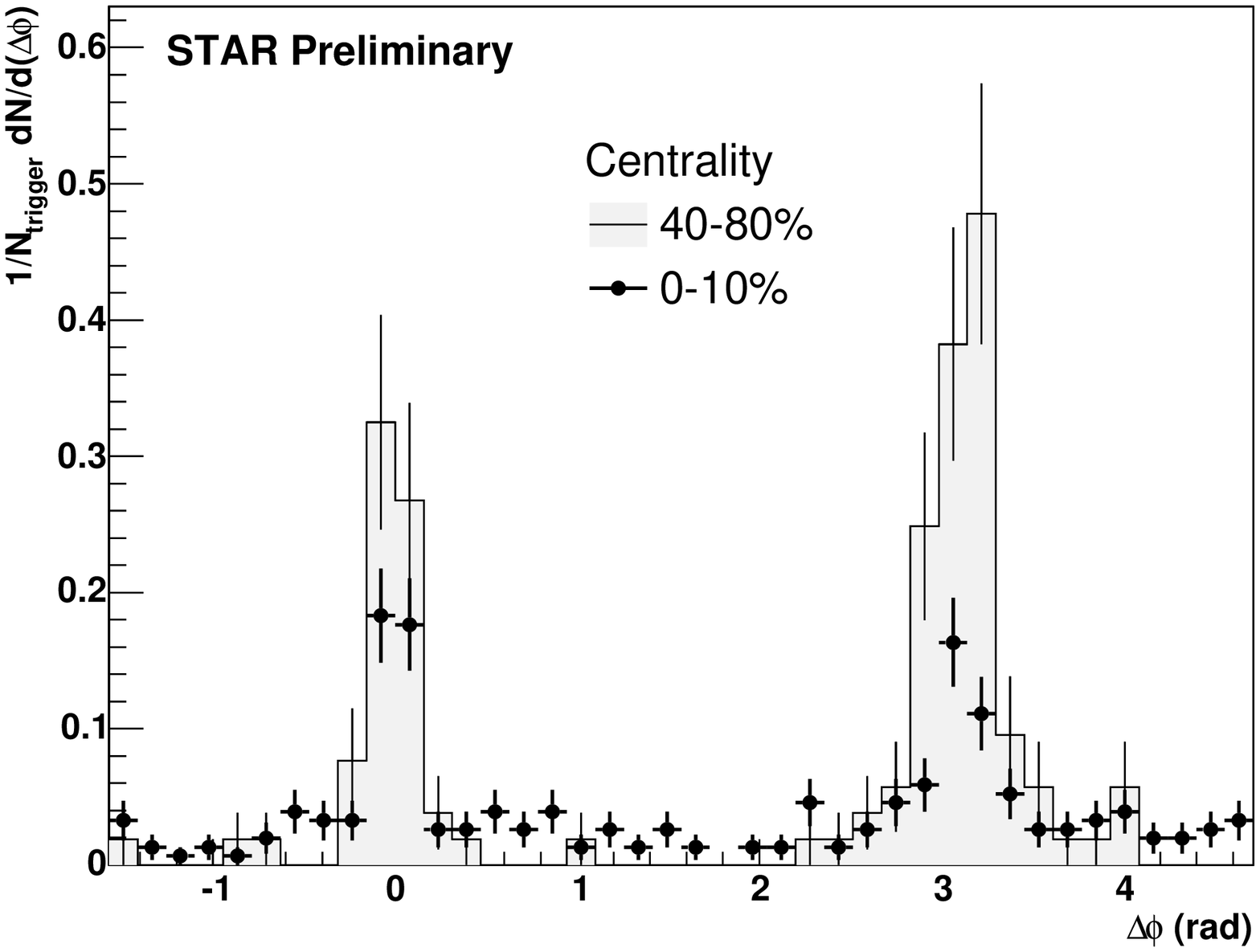}
      \vspace*{-1.7cm}
      \caption{Azimuthal correlations of trigger photons  ($ \Ettrig > 10$ GeV) with associated particles
               (4 GeV/c $ < \ptassoc < $ \Ettrig ) in central and peripheral Au+Au collisions \cite{star-dietel}.}
    \label{fig:star-dunlop6}
   \end{minipage}
   \vspace*{-0.7cm}
\end{figure}

New and additional insights are provided by extending the kinematic cuts to different \pt\ values. Two prominent examples
presented at the conference, of jet correlations at high and intermediate \pt\ values, are displayed in 
Figs.~\ref{fig:star-magestro-c} \cite{star-magestro} and \ref{fig:phenix-buesching4} \cite{phenix-buesching}, respectively.

Fig.~\ref{fig:star-magestro-c} shows jet correlations selected with high \pt\ trigger (8 $< \pttrig < 15$ GeV/c) and high \pt\ 
associated particles (6 GeV/c $ < \ptassoc < \pttrig$), in minimum bias d+Au, 20-40\% Au+Au and 0-5\% Au+Au, collisions. Very 
clear and almost background free back-to-back jets are observed in all three cases and in particular in central Au+Au, in sharp 
contrast with the away side jet disappearance in central Au+Au at lower \pt\ values ($\pttrig >4$ GeV/c and $\ptassoc >2$ GeV/c) 
previously discussed. Furthermore, while the away side yields decrease, as expected, from d+Au to central Au+Au, the near side 
yields are the same for the different systems in contrast with calculations \cite{wang-nucl-th0412061}. This new kinematic cut 
provides new constraints and challenges to theoretical models of jet quenching.

Fig.~\ref{fig:phenix-buesching4} shows the jet correlation patterns measured by PHENIX with intermediate \pt\ trigger (2.5 $< 
\pttrig < 4$ GeV) and intermediate \pt\ associated particles (2 $< \ptassoc < 3$ GeV/c) \cite{phenix-buesching}. Clear 
modifications of the away side jet as function of centrality are seen: a broad peak is observed in peripheral collisions that 
develops a dip  at $\Delta\phi = \pi$ with increasing centrality, a feature that cannot be explained with the current jet 
quenching models. Exotic features like Cherenkov \cite{cherenkov} or Mach cone \cite{casalderrey-machcone} effects are being 
discussed in this context. Measurements of three-particle correlations will be very valuable to confirm the relevance of these 
new mechanisms. Active searches are underway and first results of three-particle correlations were shown at the conference 
\cite{star-wang,phenix-ajitanand}.  

The hadron-hadron correlations discussed so far are inherently limited by the trigger bias: the energy of the jet is not well
defined by the trigger particle and the selected jets are preferentially created close to the surface. A much cleaner and
unbiased picture can be obtained in prompt photon-hadron correlations in which the hadron trigger is replaced by a photon
trigger. First attempts in this direction were presented at this conference \cite{phenix-grau,star-dietel}. An example is
displayed in Fig.~\ref{fig:star-dunlop6} showing azimuthal correlations of trigger photons  ($ \Ettrig > 10$ GeV) with 
associated particles (4 GeV/c $ < \ptassoc < $ \Ettrig ) in central and peripheral Au+Au collisions \cite{star-dietel}.

\section{Heavy Flavor}
\label{sec:heavy-flavor}

The heavy flavor results from  RHIC are among the highlights of the conference. Contrary to expectations it was shown that heavy 
quarks flow and are strongly suppressed at high \pt. These results are inferred from measurements of non-photonic electrons
which are assumed to originate from semi-leptonic decays mainly of $c$ quarks at low \pt\ but with an increasing
contribution of the heavier $b$ quarks expected to dominate the electron yield at high enough \pt.

\begin{figure}[ht]
  \vspace*{-1cm}
  \begin{minipage}[t]{70mm}
      \includegraphics*[width=7.5cm, height=6cm]{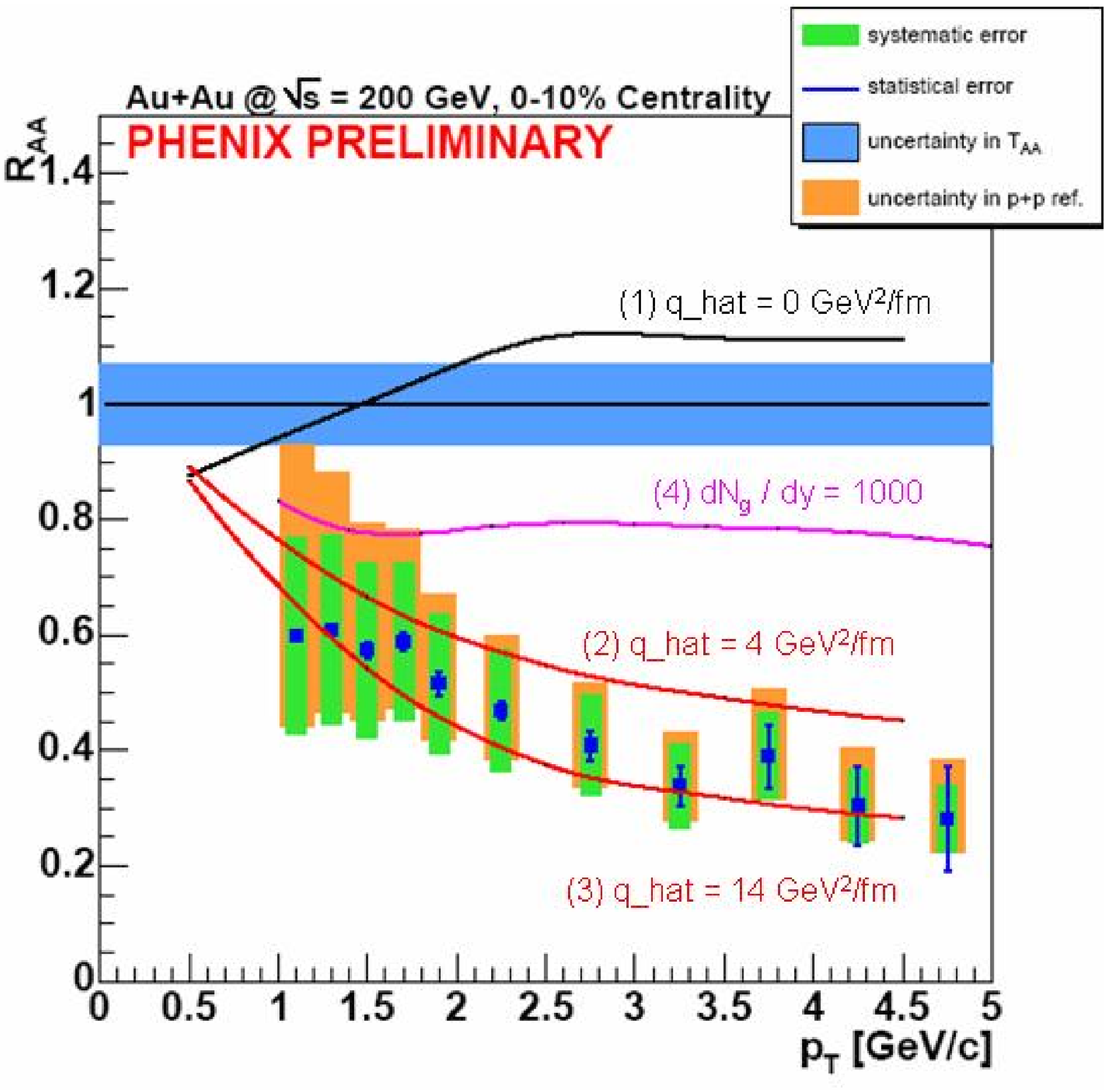}
  \end{minipage}
  \hspace{\fill}
  \begin{minipage}[t]{70mm}
     \includegraphics*[width=7.5cm, height=6.5cm]{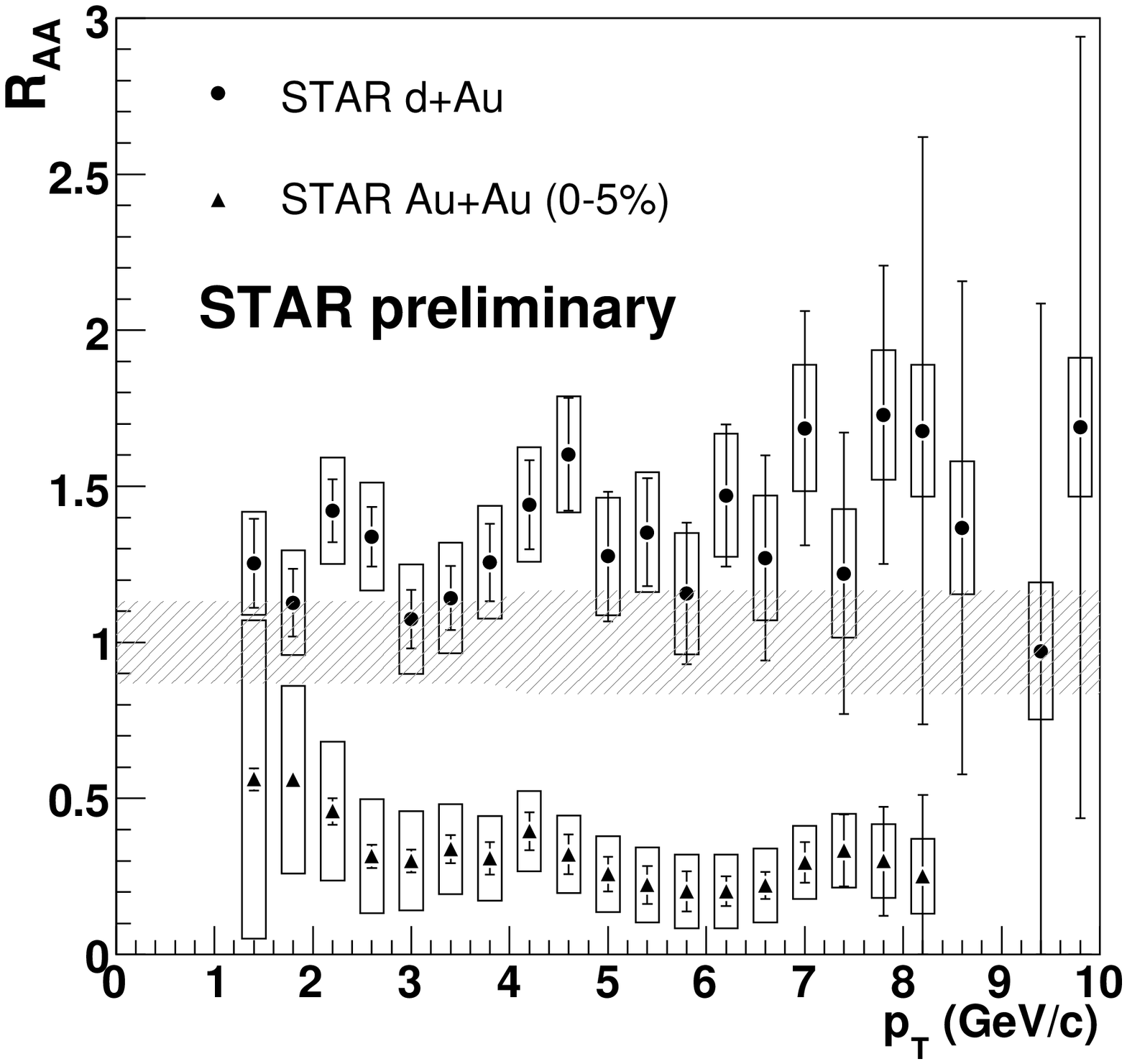}
  \end{minipage}
  \vspace*{-1cm}
  \caption{Nuclear modification factor of non-photonic electrons measured by PHENIX (left panel \cite{phenix-butsyk}) and
           STAR (right panel \cite{star-dunlop}) for central Au+Au collisions at \rootsnn = 200 GeV. Theoretical
           predictions, curves (1)-(3) from \cite{armesto05} for charm quarks only and curve (4) from \cite{djordjevic05} for charm
           and bottom quarks, are shown on the left panel.}
  \label{fig:raa-electrons-star-phenix}
  \vspace*{-0.7cm}
\end{figure}

Fig.~\ref{fig:raa-electrons-star-phenix} shows the nuclear modification factor of non-photonic electrons measured by PHENIX 
\cite{phenix-butsyk} and STAR \cite{star-dunlop} in central Au+Au collisions. The two experiments are in good agreement here. 
Both show a very strong suppression with approximately the same shape and magnitude as for hadrons (see 
Figs.~\ref{fig:phenix-shimomura1},\ref{fig:star-barannikova1}). This is a surprising result as the massive quarks are expected 
to radiate much less energy than the lighter u,d quarks.  It forces the theoretical models which incorporate contributions both 
from charm and bottom quarks, to invoke a very high initial gluon density $dN_g/dy = 3500$ \cite{djordjevic05}, a factor of 
$\sim3$ higher than for \pizero\ (see Fig.~\ref{fig:phenix-shimomura1}), in order to reproduce the observed suppression. It is a 
non-trivial challenge for the theorists to accommodate all the high \pt\ suppression results within a consistent theoretical 
framework. But there is also a not less demanding challenge for the experimentalists. The non-photonic electron results are 
derived from single electron measurements after subtracting large contributions from conversions and Dalitz decays. It is 
necessary to confirm and supplement these results with direct and separate measurements of the open charm and open bottom 
contributions. The first direct reconstruction of D mesons in Au+Au collisions reported by STAR is a first step in this 
direction \cite{star-zhang} but it seems that more substantial progress will have to await for the implementation of the vertex 
detector upgrades which are currently under development both in PHENIX and STAR.

\begin{figure}[h]
   \vspace*{-0.5cm}
   \begin{minipage}[t]{70mm}
      \includegraphics[width=7.5cm,height=5.5cm]{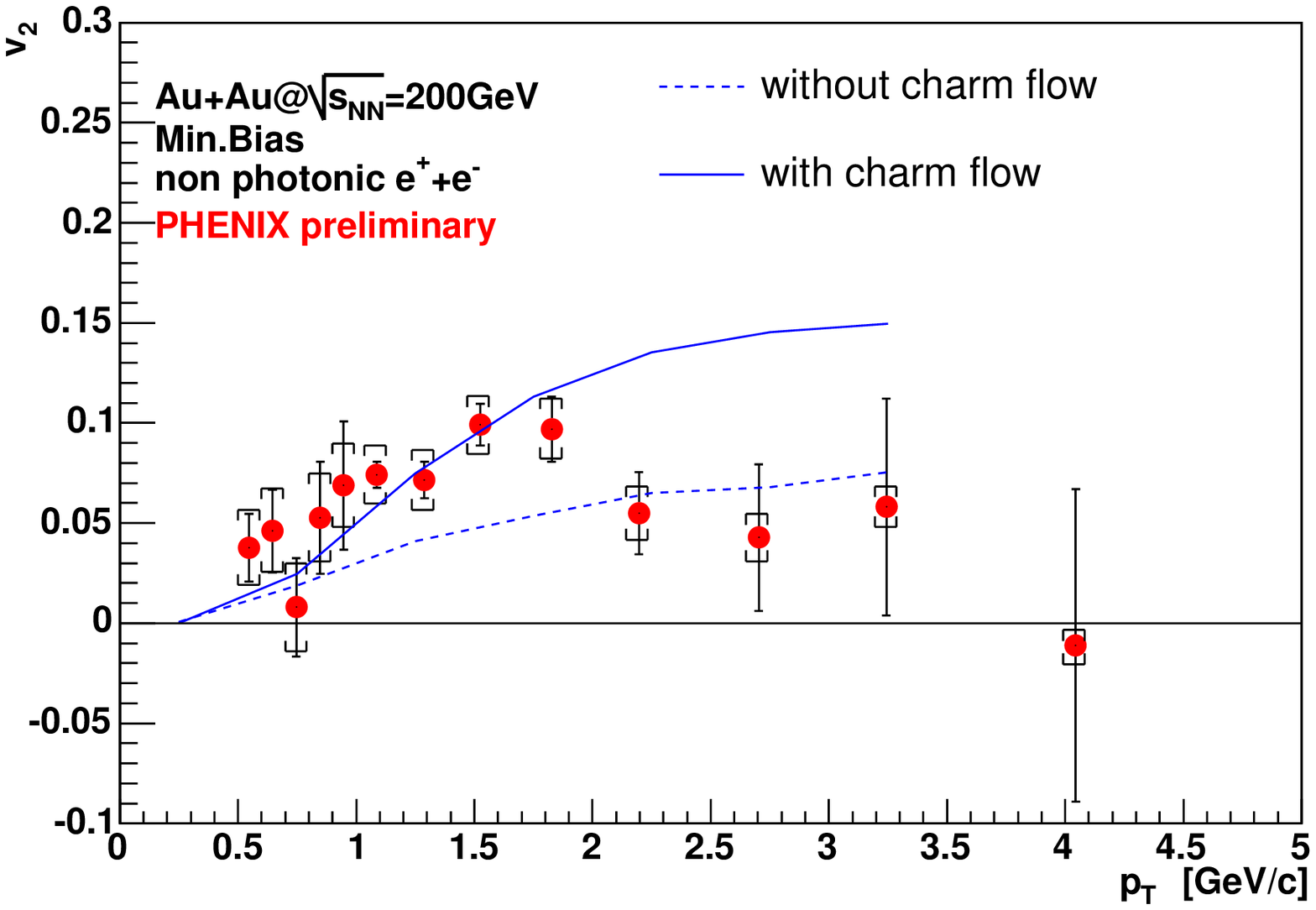}
      \vspace*{-1.7cm}
      \caption{\vtwo\ of non-photonic electrons, attributed to semi-leptonic open charm decays, measured by PHENIX 
                \cite{phenix-butsyk} compared with theoretical predictions from \cite{greco-ko-rapp}.}
    \label{fig:phenix-butsyk4}
   \end{minipage}
      \hspace{\fill}
   \begin{minipage}[t]{70mm}
      \includegraphics[width=7.5cm,height=5.5cm]{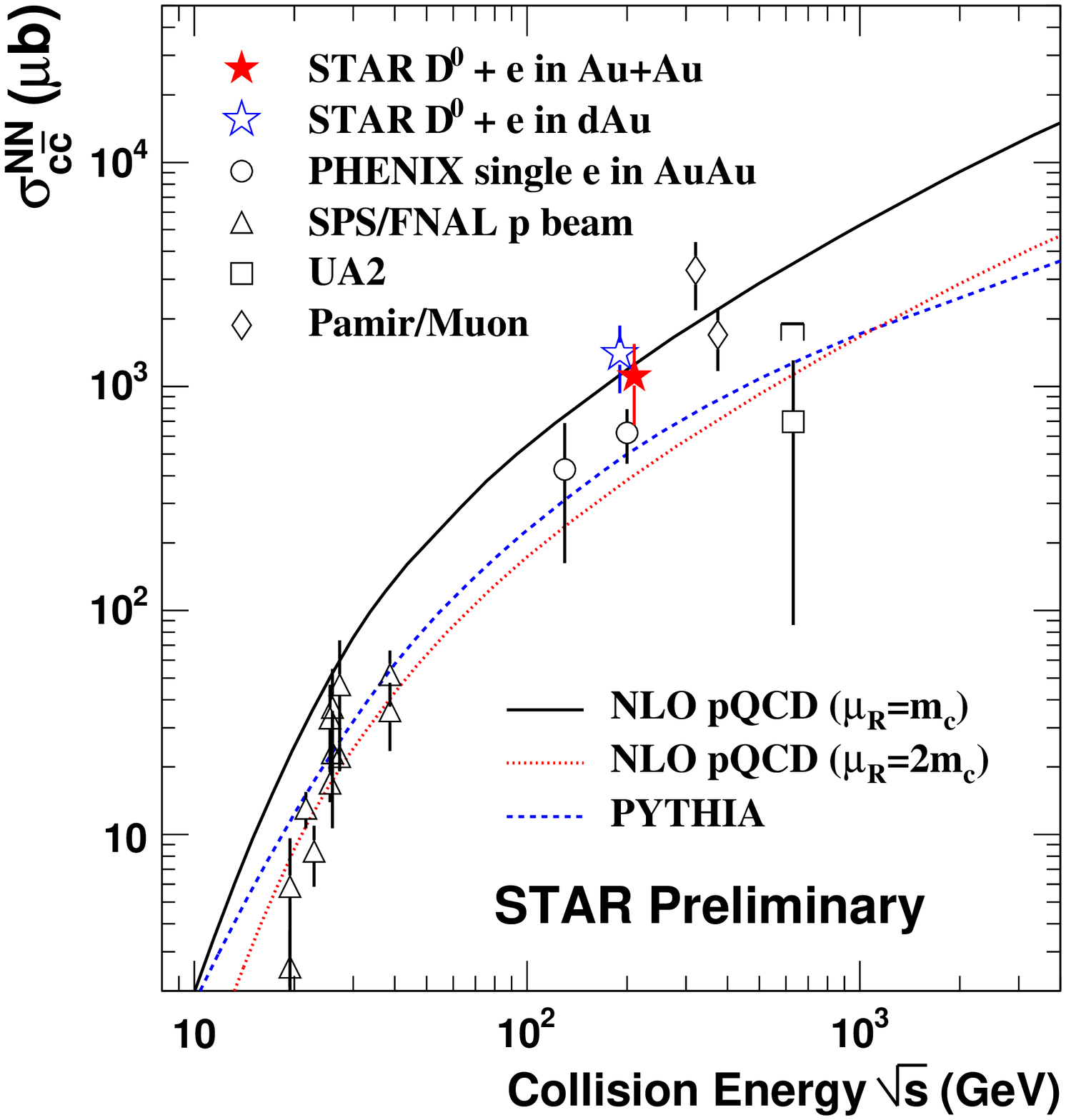}
      \vspace*{-1.7cm}
      \caption{The ${c\bar{c}}$ cross section per binary collision vs. collision energy compared to NLO pQCD
               calculations\cite{star-zhang}.}
    \label{fig:star-zhang2b}
    \end{minipage}
    \vspace*{-0.7cm}
\end{figure}

PHENIX reported measurements of elliptic flow \vtwo\ of non-photonic electrons as function of \pt\ \cite{phenix-butsyk}. The
results are shown in Fig.~\ref{fig:phenix-butsyk4}. The sizable flow observed at \pt values of 1-1.5~GeV/c, where the yield is 
dominated by semi-leptonic decays of open charm, indicates a sizable flow of D mesons. The flow seems to decrease to vanishing 
values at high \pt\ (the statistical errors are too large for a more conclusive statement) where the contribution from 
semi-leptonic bottom decays could be significant. A comparison to calculations with and without charm flow \cite{greco-ko-rapp} 
favors the interpretation of charm flow. This implies a strong interaction with the medium and a high degree of thermalization 
of the charm quarks, again reinforcing the case for a strongly coupled QGP. Direct and separate measurements of D and B mesons 
elliptic flow will be very valuable to firmly and unambiguously establish the non-photonic electron results.
 
PHENIX has translated the non-photonic electron measurements in p+p and minimum bias Au+Au collisions at \rootsnn\ = 200 GeV 
into a total cross-section of charm production per nucleon-nucleon collision of $\sigma_{c\bar{c}}^{NN}$ = 
0.92$\pm$0.15(stat.)$\pm$0.54(sys.) mb \cite{phenix-charm-pp}  and $\sigma_{c\bar{c}}^{NN}$ = 
0.62$\pm$0.057(stat.)$\pm$0.16(sys.) mb \cite{phenix-charm-auau}, respectively. STAR has done a similar thing from the combined 
electron and direct D meson data in minimum bias d+Au ($\sigma_{c\bar{c}}^{NN}$ = 1.4$\pm$0.2(stat.)$\pm$0.2(sys.) mb) and Au+Au 
($\sigma_{c\bar{c}}^{NN}$ = 1.11$\pm$0.08(stat.)$\pm$0.42(sys.) mb) measurements \cite{star-zhang}. The PHENIX and STAR results 
are consistent within their large error bars and they support the scaling of the charm production cross section with the number 
of binary collisions. However, there is a factor of almost two difference in their minimum bias Au+Au cross sections. Larger 
precision is needed in view of the importance of this cross section in the evaluation of the \Jpsi\ production through charm 
recombination (see Section~\ref{sec:Jpsi}) and to better constrain the NLO pQCD calculations. See Fig.~\ref{fig:star-zhang2b} 
where the STAR and PHENIX results are plotted in a compilation of cross section vs. collision energy and compared to NLO pQCD 
calculations \cite{star-zhang}.

NA60 shed light on the long standing question of the origin of the dimuon excess at intermediate masses (m= 1.2 - 2.7 GeV/c$^2$)
observed by NA50 in S+U and Pb+Pb collisions \cite{na50-int-mass}. This excess can be interpreted as an enhancement of charm 
production in nuclear collisions as argued by NA50, or as evidence for thermal radiation \cite{rapp-shuryak,li-gale}. NA60 
confirmed that also in In+In collisions the dimuon yield at intermediate masses is enhanced over the expected contributions from 
Drell-Yan (fixed by the high mass region) and open charm (assuming $\sigma_{c\overline{c}}$=12~$\mu$b/nucleon), in very good 
agreement with the previous NA50 observations. But using displaced vertex information to tag the open charm semileptonic decays, 
NA60 was able to go one step further and to prove that the excess originates from a prompt source of dimuons rather than an 
enhanced open charm yield \cite{na60-scomparin}. If it needed any proof, these results clearly illustrate the superiority of a 
direct measurement or tagging of D-mesons over more inclusive measurements.

\section{J/$\psi$}
\label{sec:Jpsi}

   The \Jpsi\ production keeps generating great interest and remains an exciting topic. We have seen updated results from NA50,
new results from NA60 and last but not least first results from PHENIX at RHIC. All these results taken together with the recent
lattice calculations which predict the \Jpsi\ to survive as a bound state in the QGP up to temperatures of $\sim 1.5-2 T_c$, 
call for redefining the significance of the \Jpsi\ production as \emph {the} diagnostic tool of deconfinement at SPS and RHIC 
energies.

\begin{figure}[h]
   \vspace*{-0.7cm}
   \centering
   \includegraphics*[width=7.5cm, height=6.5cm]{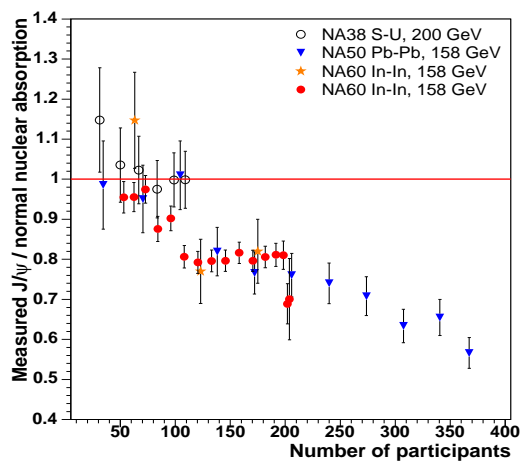}
   \vspace*{-1cm}
   \caption{\Jpsi\ suppression pattern measured in \mbox{S+U}, and \mbox{Pb+Pb} by NA50 and in \mbox{In+In} by NA60, as a
            function of N$_{part}$ \cite{na60-scomparin}.}
   \label{fig:na60-scomparin14}
   \vspace*{-0.7cm}
\end{figure}

With a new determination of the \Jpsi\ nuclear absorption cross section, derived from precision pA data, NA50 confirmed that the 
\Jpsi\ yield in S+U and peripheral Pb+Pb collisions is consistent with normal nuclear absorption. Only semi-central or central 
Pb+Pb collisions show an additional suppression (see Fig.~\ref{fig:na60-scomparin14}) \cite{na50-ramello}. NA50 further 
characterized this anomalous suppression with two new results: (i)the suppression occurs mainly at low \pt. (ii) the central to 
peripheral ratios \RCP\ as function of \pt\ show an initial increase with \pt\ followed by a saturation at the level of 1 at 
higher \pt\ demonstrating binary scaling (see Fig.~12 of \cite{na50-ramello}).

The first NA60 results on In+In collisions show also an anomalous \Jpsi\ suppression. The pattern is very similar to the one
observed by NA50 as illustrated in Fig.~\ref{fig:na60-scomparin14} where both results are overlayed together
\cite{na60-scomparin}. Within the accuracy of the measurements, the figure supports the claim already made in
Section~\ref{sec:system-size} that the  number of participants seems to be a good scaling factor to take account of system size
effects.

\begin{figure}[h]
    \vspace*{-5mm}
    \begin{center}
    \begin{tabular}{cc}
    \includegraphics[width=75mm,clip=true]{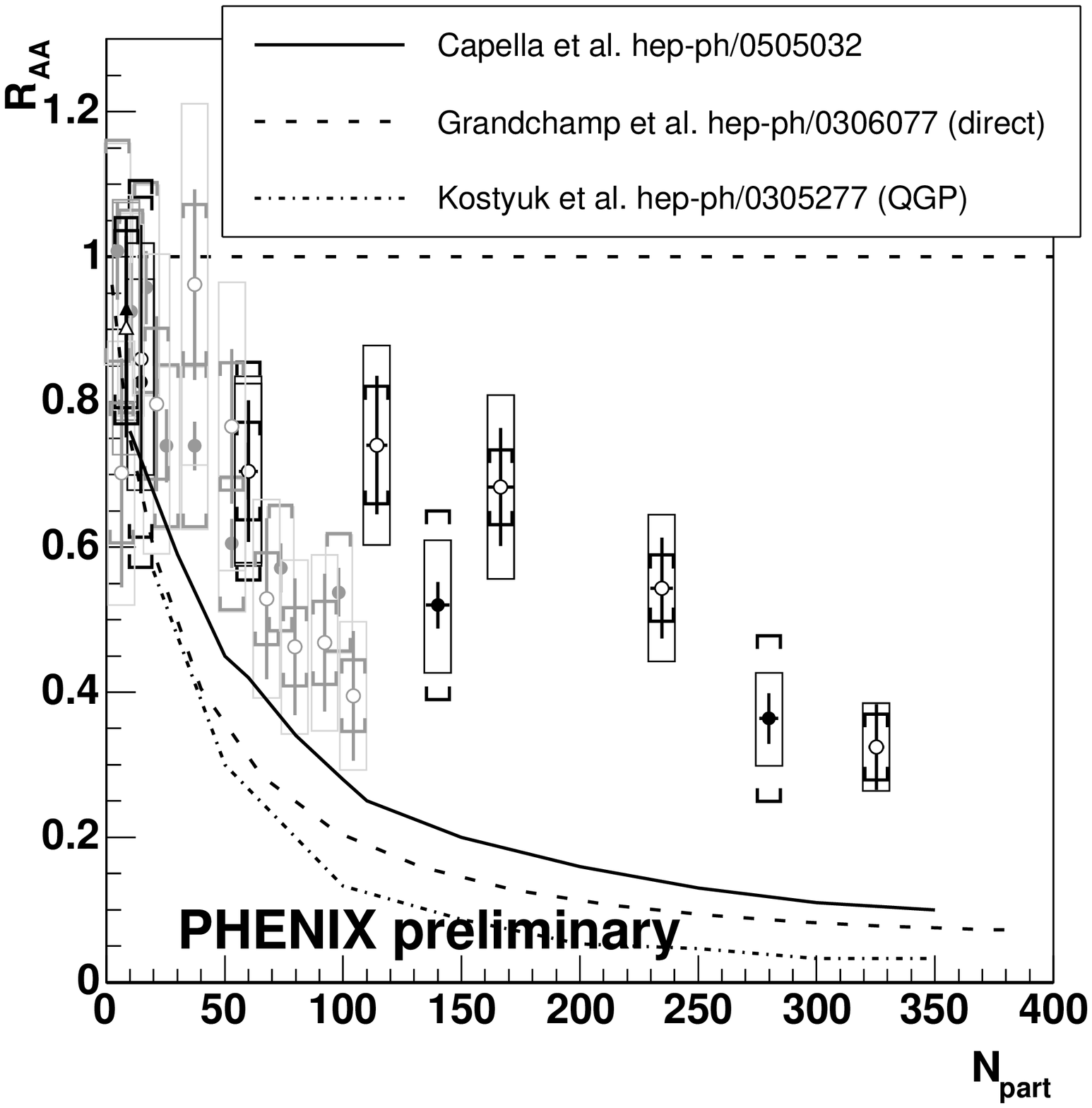}&
    \includegraphics[width=75mm,clip=true]{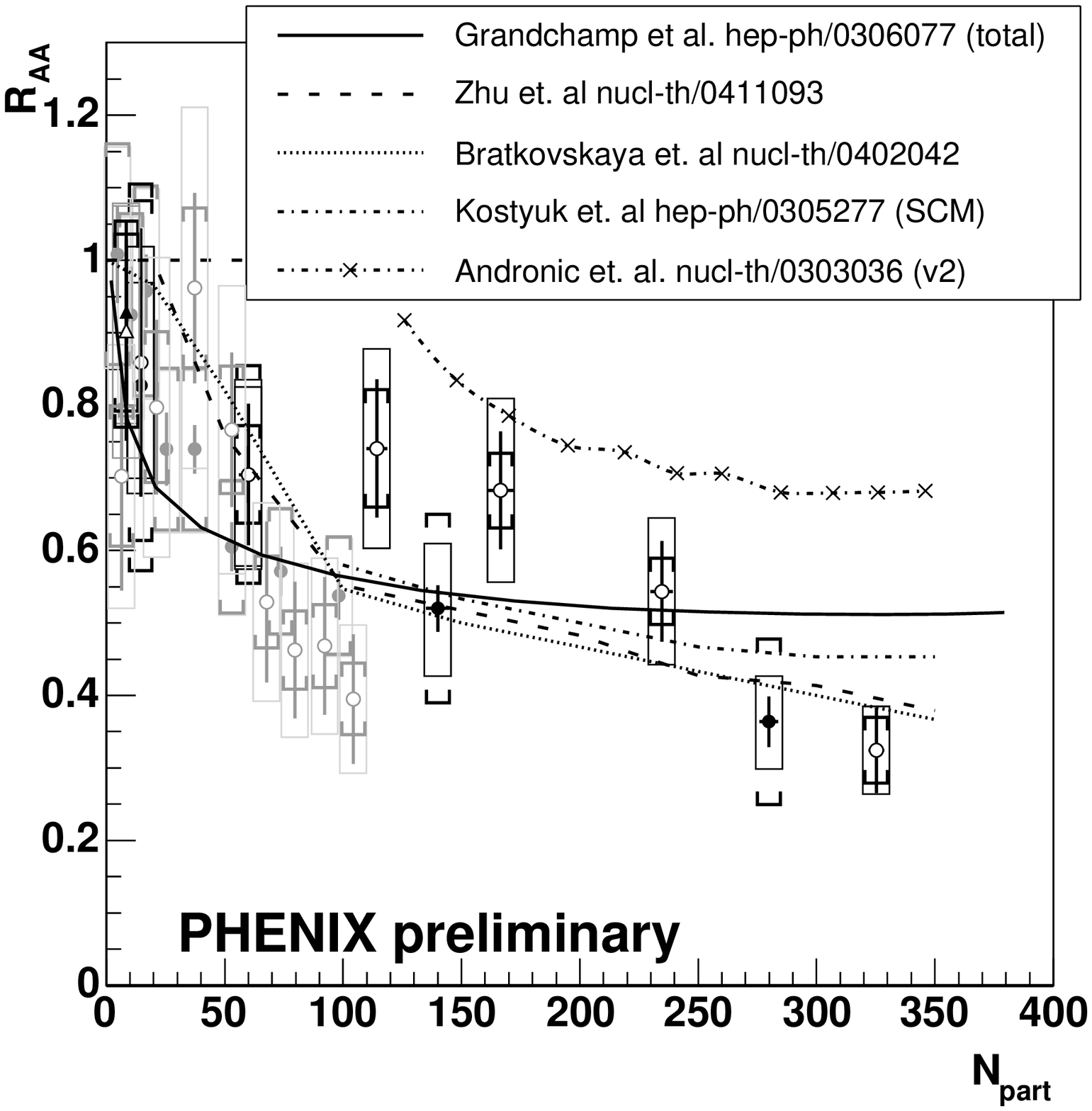}\\
    \end{tabular}
    \end{center}
    \vspace*{-15mm}
    \caption{\Jpsi\ nuclear modification factor as a function of the number of participants in d+Au, Au+Au and Cu+Cu measured
              by PHENIX at
             \rootsnn\ = 200 GeV, compared to models which were able to explain the NA50 anomalous suppression  (left panel)
             \cite{capella,kostyuk,grandchamp} and models involving either \Jpsi\ regeneration by quark recombination
             \cite{kostyuk,grandchamp,bratkovskaya,andronic} or \Jpsi\ transport in medium \cite{zhu} (right panel)
              \cite{phenix-buesching}.}
    \vspace*{-0.6cm}
    \label{fig:phenix-buesching3}
\end{figure}

PHENIX presented results  from a comprehensive set of \Jpsi\ measurements. The set includes \Jpsi\ production in p+p, d+Au, 
Cu+Cu and Au+Au collisions at \rootsnn\ =200 GeV and Cu+Cu at 62 GeV, at mid-rapidity ($|\eta| < $ 0.35) through the \ee\ decay 
channel and at forward rapidity ($|\eta|\in[1.2,2.2]$) through the \mumu\ decay channel \cite{phenix-buesching}. The nuclear 
modification factor \RAA\ vs. centrality is shown in Fig.~\ref{fig:phenix-buesching3} for the \rootsnn = 200 GeV data. There is 
a clear suppression and within the present errors, which are still too large, the PHENIX data show the same trend and the same 
magnitude of suppression irrespective of species and energy. Furthermore, the magnitude of the suppression, reaching a factor of 
$\sim$3 for the most central collisions, is similar to the one observed at the SPS.

This universality of the suppression is a surprising result. Models based on interactions with comovers \cite{capella}, color
screening \cite{kostyuk}, or QCD-inspired in-medium effects \cite{grandchamp}, which were able to explain the anomalous \Jpsi\
suppression at the SPS, do predict a stronger suppression at the RHIC as one would intuitively expect due to the higher 
densities of RHIC (see left panel of Fig.~\ref{fig:phenix-buesching3}).

The discrepancy is resolved by invoking the regeneration of \Jpsi\ at a later stage of the collision via recombination of $c$ 
and $\bar{c}$ quarks, more abundantly produced at RHIC. Several attempts in this direction 
\cite{kostyuk,grandchamp,bratkovskaya,andronic}, combining suppression and recombination, do reproduce the data reasonably well 
right panel of Fig.~\ref{fig:phenix-buesching3} \footnote{the figure includes also calculations based on \Jpsi\ transport in the 
QGP \cite{zhu} which are also in reasonable agreement with the data.}. This appears as an additional score in favor of the 
recombination models, whose success in explaining other aspects of the data, like e.g. the flow pattern at intermediate \pt\, 
was already emphasized. Recombination has however, additional consequences on \Jpsi\ properties which should be corroborated by 
the data before it can be accepted as the dominant mechanism for J/$\psi$ production at RHIC. First, the \Jpsi\ rapidity 
distribution is expected to become narrower with increasing centrality, a feature not supported by existing results 
\cite{phenix-pereira}. (This negative observation should be taken with some reservation since it is based on results suffering 
from large uncertainties). Second, the \Jpsi\ should exhibit a high \pt\ suppression pattern compatible with the one observed 
for charm quarks (see Section~\ref{sec:heavy-flavor}). Finally the \Jpsi\ should exhibit elliptic flow consistent with the flow 
observed for charm quarks. Results on this crucial test are not yet available. For a better scrutiny of the recombination models 
we thus need more precise data on the rapidity distribution and  high \pt\ suppression of the \Jpsi\ and equally precise data on 
elliptic flow. A more accurate determination of the charm production cross section is also essential. On the theory side, firm 
predictions of the correlations between the properties of $c$ quarks and \Jpsi\ are also needed.

 Another interesting possibility, appealing for its simplicity, has been proposed that could explain all
the SPS and RHIC results. The anomalous \Jpsi\ suppression would originate from the total melting of excited charmonium states,
in particular the $\chi_c$, that feed the \Jpsi\ and account for about 40\% of its yield \cite{nardi}. This is supported by
recent lattice QCD calculations according to which, at the temperatures reached at SPS and RHIC energies, only the $\psi$' and
$\chi_c$ are dissolved by color screening in the QGP, whereas the \Jpsi\ remains as a bound state up to temperatures of $\sim
1.5-2 T_c$ \cite{asakawa,datta}. If this explanation holds true, we will have to wait to the LHC, where the temperature will be
high enough, to observe the suppression of directly produced \Jpsi.

\section{Low-mass dileptons}
\label{sec:low-mass}

Many new and significant results have been presented at this conference: first attempts of PHENIX to measure low-mass electron 
pairs from the 2004 high luminosity Au+Au run, almost final results of CERES from the 2000 Pb+Au run and last but not least, the 
superb data of NA60 on In+In collisions.

\begin{figure}[h]
    \vspace*{-0.3cm}
    \begin{minipage}[l]{0.5\textwidth}
    \includegraphics*[width=7.5cm, height=6.0cm]{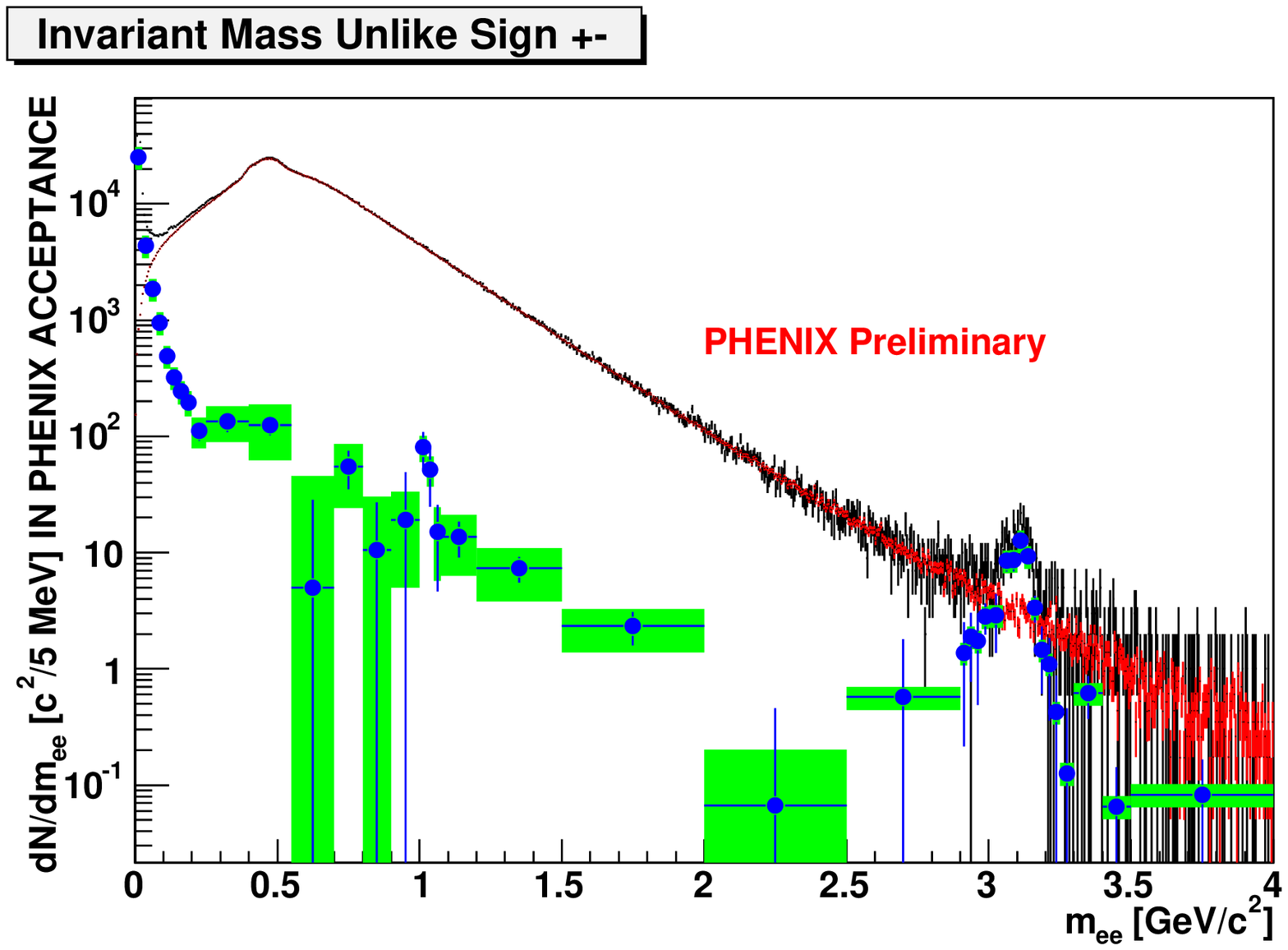}
    \end{minipage}
    \hfill
    \begin{minipage}[l]{0.5\textwidth}
    \includegraphics*[width=7.5cm, height=6.0cm]{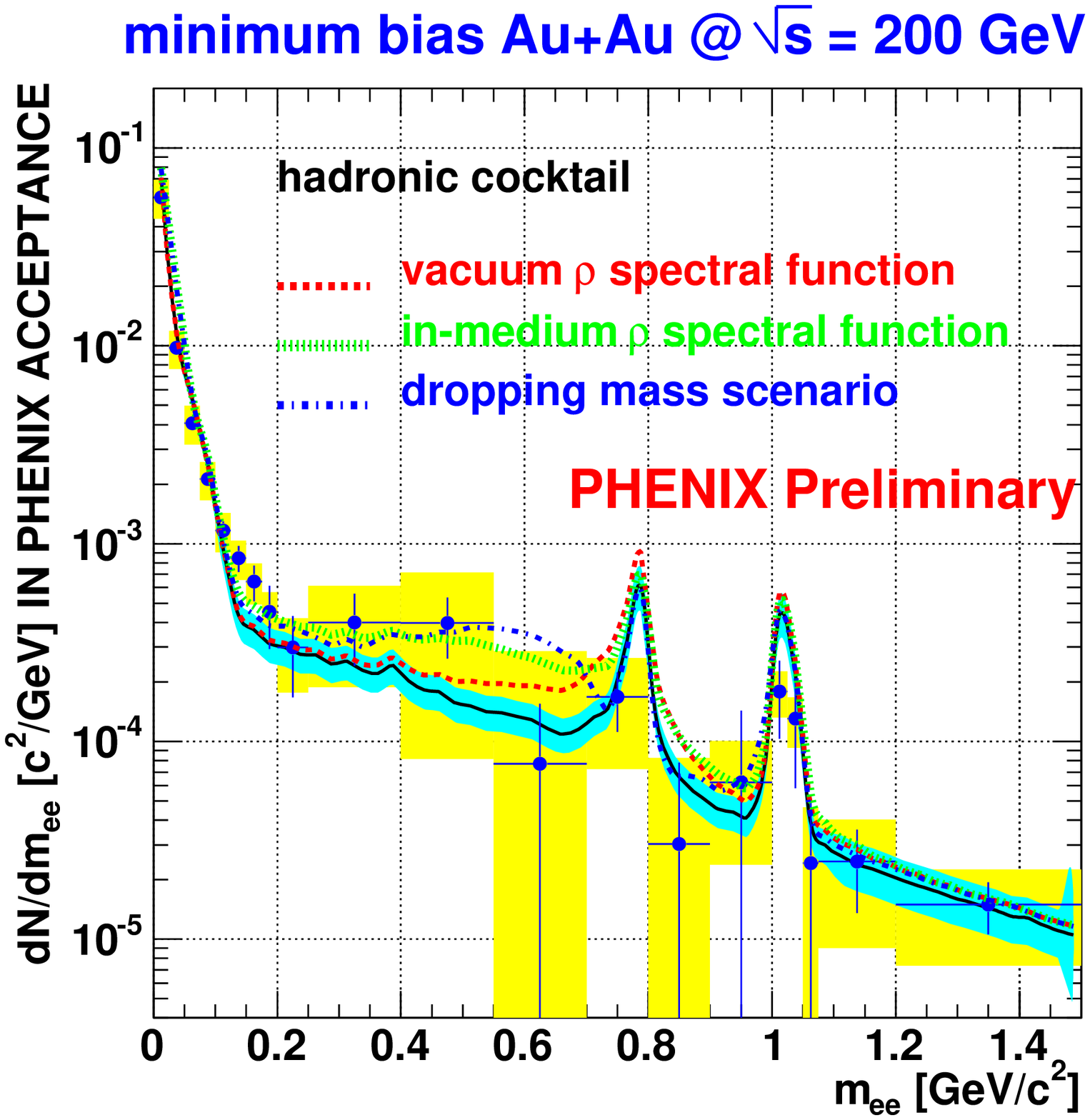}
    \end{minipage}
    \vspace*{-10mm}
    \caption{Low-mass dilepton results of PHENIX in Au+Au collisions at \rootsnn\ = 200 GeV. Left panel: measured unlike sign
            (black), mixed events background (red), and subtracted (blue), mass spectra. Right panel: subtracted mass spectrum
            compared to model predictions \cite{rapp} and to a pure hadronic cocktail \cite{phenix-alberica}}
    \label{fig:phenix-alberica}
    \vspace*{-0.7cm}
\end{figure}

PHENIX has an excellent mass resolution (approximately 1\% at the $\phi$ meson mass) and good electron identification
capabilities in the central arms, based on a RICH detector and an electro-magnetic calorimeter. However, the limited azimuthal
acceptance in the central arms and the strong magnetic radial field beginning at the vertex, make the identification and
rejection of the relatively large number of electron-positron pairs from Dalitz decays and photon conversions very difficult,
resulting in an overwhelming combinatorial background. In the present set-up the signal to background S/B ratio for masses m = 
0.2 - 0.5 GeV/c$^2$ is smaller than 1/100 as illustrated in the left panel of Fig.~\ref{fig:phenix-alberica} 
\cite{phenix-alberica}.
PHENIX has mastered the background subtraction technique to a very high level of precision ($\pm$0.25\%). But with a S/B $<$ 
1/100 this is not sufficient. The statistical significance of the measurement is largely reduced and the systematic uncertainty 
is large preventing any conclusive interpretation of the data. The right panel of Fig.~\ref{fig:phenix-alberica} shows the 
background subtracted spectrum compared to a pure cocktail of expected sources (hadronic decays and semi-leptonic open charm 
decay) as well as several model predictions \cite{rapp} including the vacuum $\rho$ spectral function, the in-medium $\rho$ 
broadening and the in-medium $\rho$ dropping mass. Differences as large as a factor of three between the various predictions 
cannot be resolved within the present uncertainties of the data. PHENIX has also the potential of measuring the $\phi$ meson 
through the \ee\ and $K^+K^-$ decay channels. Although the S/B ratio in the $\phi \rightarrow $ \ee\ is somewhat better (S/B 
$\sim$ 1/60) the errors are still too large for a meaningful comparison to the much higher precision of the $\phi \rightarrow 
K^+K^-$ data \cite{phenix-kozlov}. Larger data samples will be of very limited help. A breakthrough in the PHENIX capabilities 
to measure low-mass dileptons is expected with the installation foreseen in 2006, of the Hadron Blind Detector \cite{hbd-nim} 
presently under construction \cite{phenix-ravinovich}.

\begin{figure}[h]
    \centering
    \includegraphics*[width=7.5cm, height=6.0cm]{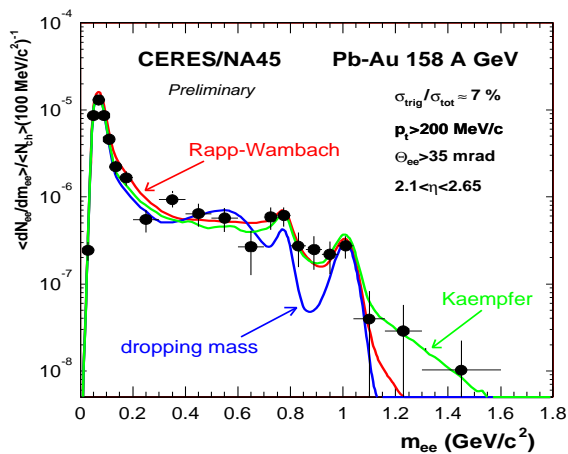}
    \vspace*{-10mm}
    \caption{Dielectron mass spectrum measured by CERES compared to calculations performed by Rapp (curves labeled Rapp-Wambach
             and dropping mass) \cite{rapp-ceres} and by Kaempfer \cite{kaempfer} (see text for further details)
             \cite{ceres-miskowiec}.}
   \label{fig:ceres-miskowiec4b}
   \vspace*{-1.0cm}
\end{figure}

CERES presented almost final results of the 2000 Pb+Au run with the upgraded spectrometer with a radial TPC. The absolutely
normalized invariant mass spectrum of electron pairs shows a clear excess with respect to the hadron cocktail of expected
sources, at masses m $>$ 0.2 GeV/c$^2$ (see Fig. 4 in ref. \cite{ceres-miskowiec}). The magnitude of the excess  and its \pt\
dependence are in very good agreement with previous CERES results. With the improved mass resolution a hint of the $\omega$ and
$\phi$ meson peaks is seen for the first time in the CERES data. The results are compared in Fig.~\ref{fig:ceres-miskowiec4b} to 
calculations performed by Rapp \cite{rapp-ceres} including in-medium modifications of the $\rho$ meson (dropping mass according 
to the Brown-Rho scaling \cite{brown-rho}, broadening of the $\rho$ according to the Rapp-Wambach model \cite{rapp-wambach}). 
The figure also shows a calculation by Kaempfer based on a parametrization of the dilepton yield in terms of $q\bar q$ 
annihilation inspired by quark-hadron duality \cite{kaempfer}. Whereas the three calculations give very similar results for 
masses m $<$ 0.8 GeV/c$^2$ where the precision of the data is insufficient to discriminate between them, it is interesting to 
note that the pronounced minimum between the $\omega$ and the $\phi$ in the dropping mass scenario, as implemented in the Rapp 
calculations, is not seen in the data.

\begin{figure}[h]
    \vspace*{-0.8cm}
    \begin{minipage}[l]{0.5\textwidth}
    \includegraphics*[width=7.5cm, height=6.0cm]{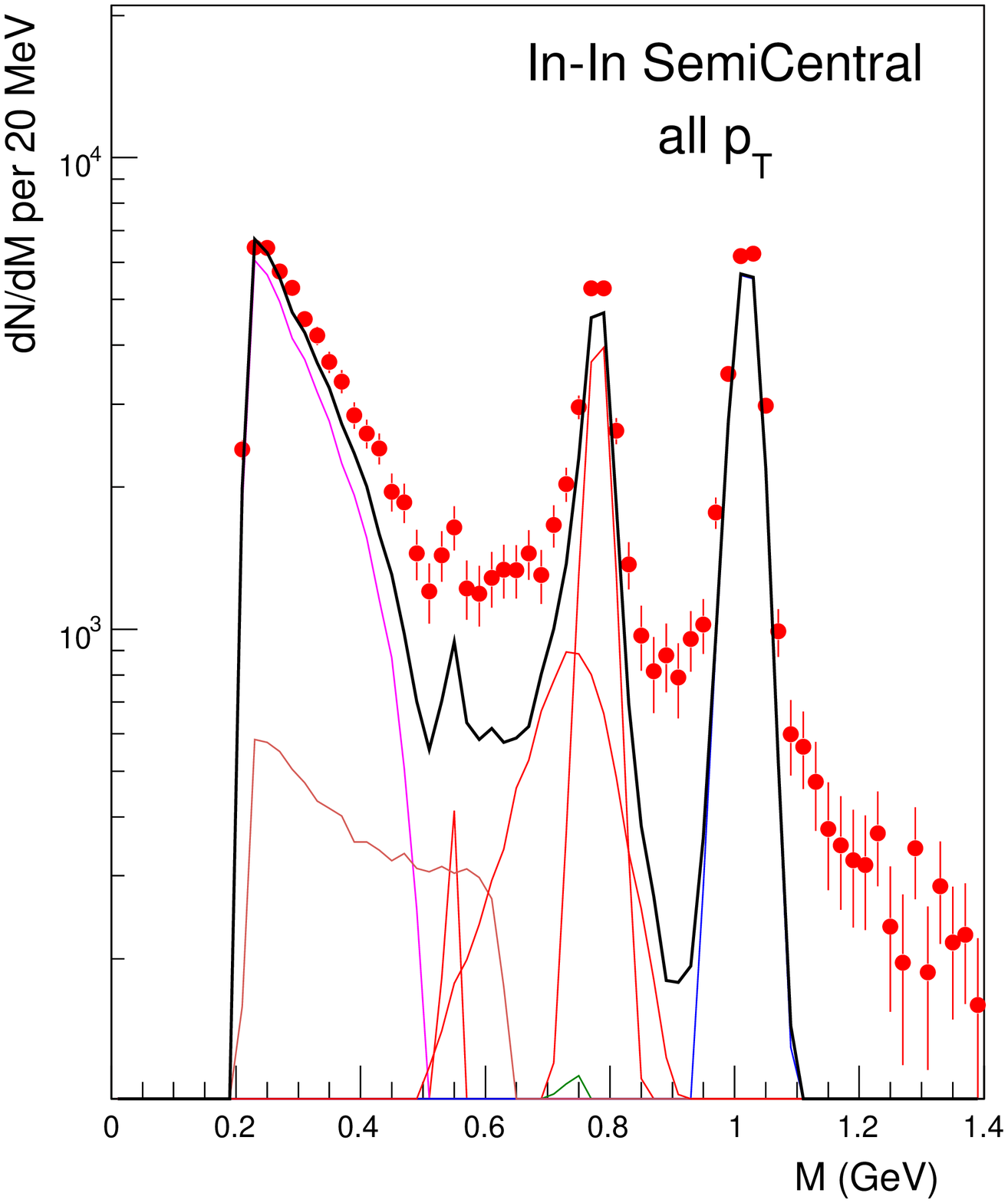}
    \end{minipage}
    \hfill
    \begin{minipage}[l]{0.5\textwidth}
    \includegraphics*[width=7.5cm, height=6.0cm]{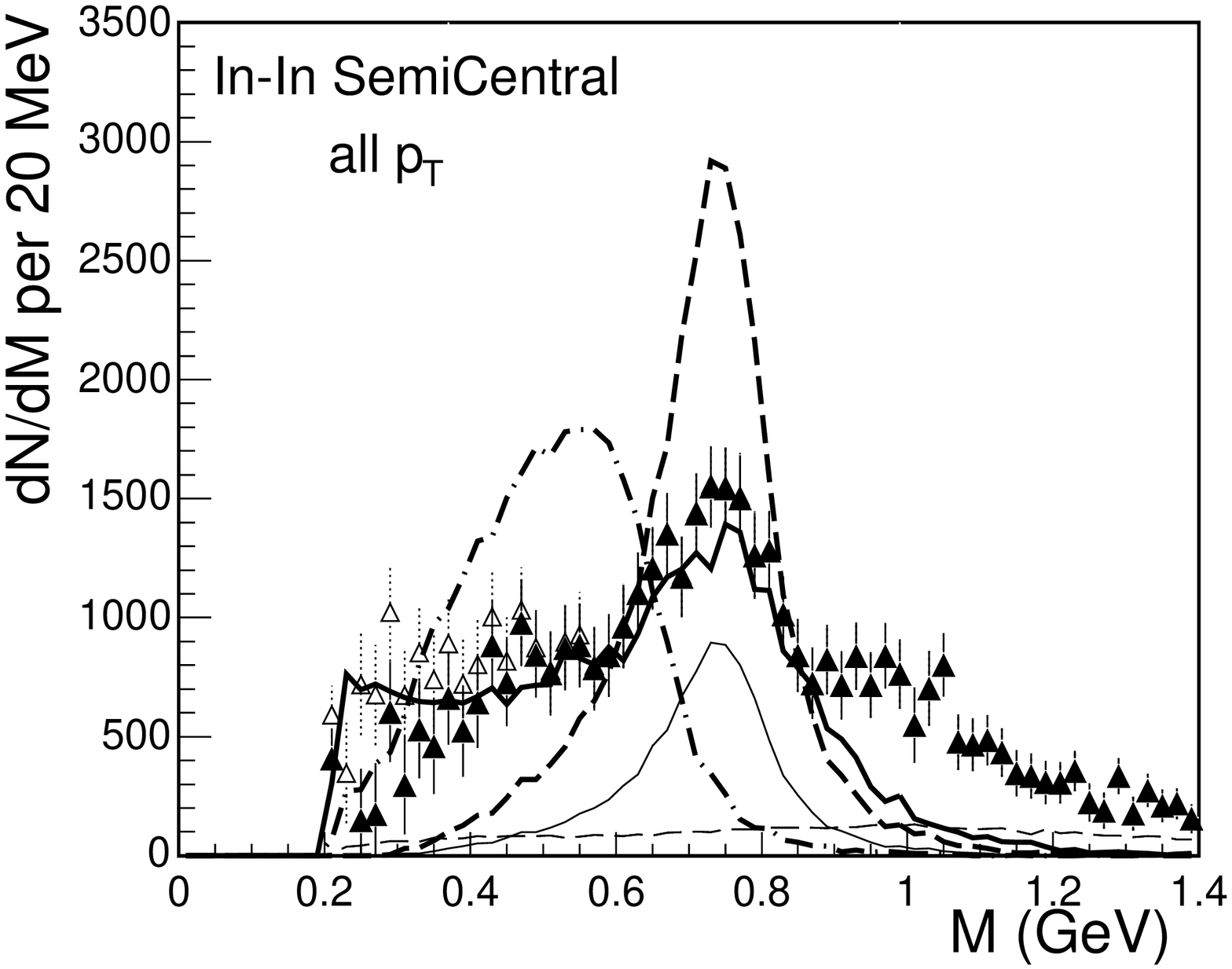}
    \end{minipage}
    \vspace*{-10mm}
    \caption{NA60 dimuon results in semi-central In+In collisions. Left panel: mass spectrum compared to hadronic cocktail.
             Right panel: excess mass spectrum compared to the cocktail $\rho$ (thin solid line), open charm decays
             (thin dashed line), and calculations \cite{rapp-na60} including the vacuum shape (thick dashed line), in-medium broadening
             (thick solid line) and dropping mass (dashed dotted line), of the $\rho$
              \cite{na60-sanja}.}
    \label{fig:na60-scomparin8}
    \vspace*{-0.9cm}
\end{figure}

The highlight on low-mass dileptons belongs without any doubt to NA60. With an excellent mass resolution of 2.2\% at the $\phi$ 
mass and excellent statistics, NA60 presented high quality results on low-mass dimuons in 158 AGeV In+In collisions 
\cite{na60-scomparin,na60-sanja}. The data show a clear excess of dimuons which increases with centrality and is more pronounced 
at low pair \pt. The results confirm, and are consistent (at least qualitatively) with, the pioneering CERES results over the 
last ten years. NA60 also showed (see Fig.~\ref{fig:na60-scomparin8}) the excess mass spectrum, obtained after subtracting from 
the data the hadronic cocktail without the $\rho$ (see \cite{na60-sanja} for details of the subtraction procedure). The excess 
exhibits a  broad structure, with a width increasing with centrality, centered at the nominal position of the $\rho$ meson mass.
The figure shows also a comparison of the excess with model calculations \cite{rapp-na60} including the vacuum $\rho$ (thick
dashed line), in-medium $\rho$ broadening (thick solid line) according to the Rapp-Wambach model \cite{rapp-wambach} and 
dropping $\rho$ mass (dashed-dotted line). The comparison clearly favors the in-medium broadening of the $\rho$ meson over the 
dropping mass scenario as implemented in \cite{rapp-na60}. The conclusions are valid also for other centralities and as function 
of \pt.

The NA60 data certainly pose a new constraint on the theoretical models which will have to simultaneously account for the CERES 
and NA60 results. But the deeper impact of these results is their possible relevance to the broader context of chiral symmetry 
restoration. If the system reaches, or is near to, chiral symmetry restoration then the dilepton results could be telling us 
that the approach to such a state proceeds through broadening and eventually subsequent melting of the resonances rather than by 
dropping masses or mass degeneracy between chiral partners.

\section{Photons}
\label{sec:photons}

\begin{figure}[ht!]
    \centering
    \vspace*{-5mm}
    \includegraphics[height=6.0cm]{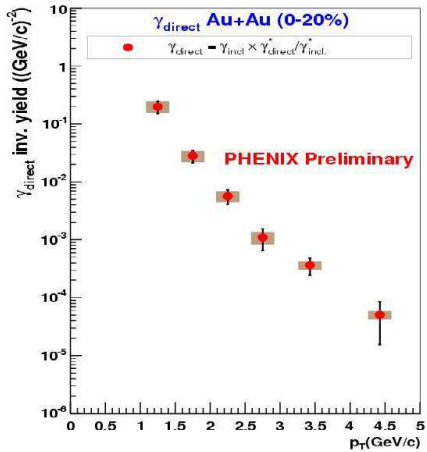}
    \vspace*{-10mm}
    \caption{ \pt\ distribution of direct photons measured by PHENIX in central Au+Au collisions at  \rootsnn\ = 200 GeV
              \cite{phenix-akiba}. }
    \label{fig:akiba4-photon}
    \vspace*{-10mm}
\end{figure}

Using a novel analysis technique based on low-mass electron pairs with high \pt, PHENIX presented interesting and intriguing
results on direct photons. Exploiting the fact that any source of real $\gamma$ photons emits also virtual $\gamma^*$ photons
with very low-mass, the low-mass electron pair yield (after taking into account the contribution from Dalitz decays) is
translated into a direct photon spectrum assuming $\gamma_{direct}/\gamma_{incl.} = \gamma^{*}_{direct}/\gamma^{*}_{incl.}$
\cite{phenix-bathe,phenix-akiba}. The resulting direct photon spectrum is shown in Fig.~\ref{fig:akiba4-photon}. It is 
compatible with the spectrum obtained from a conventional analysis of inclusive real photons \cite{phenix-photons} but it has 
smaller error bars giving more significance to the signal and allowing to extend its range down to \pt = 1 GeV/c. These two 
improvements are crucial in making apparent an excess of direct photons in the \pt\ range of 1 to 4 GeV/c, over NLO pQCD 
calculations \cite{vogelsang}, which is then ascribed to emission from the medium. Interpreted as the elusive thermal radiation 
of the QGP, the excess spectrum implies an initial temperature of 570 MeV or an average temperature of 360 MeV 
\cite{d'enterria}, in the plasma.
But is this a unique interpretation? Are these direct photons of thermal origin? More studies, in particular applying the same
analysis technique to reference measurements of p+p and d+Au, are needed to consolidate the findings.

\section{Outlook}
\label{sec:outlook}
 To summarize the summary, this was a great conference.
I tried to focus here on the highlights among the many new and exciting results that were presented. In the "Focus" talks, STAR 
summarized its results \cite{star-gagliardi} and PHENIX synthesized its findings \cite{phenix-akiba} drawing the emerging 
picture that characterizes the matter formed at RHIC. A very high density, strongly interacting partonic matter (made of 
constituent quarks) is formed, reaching thermal equilibrium very early in the collision and behaving like a perfect fluid with a 
very low (zero?) viscosity. This is not the ideal QGP gas of free quarks and gluons that we have been looking for during almost 
twenty years. This discovery is still ahead of us. The higher energies of the LHC might be needed for that. As emphasized in the 
Introduction, RHIC has accomplished a lot but much is left to do. We are only starting to examine the detailed properties of the 
matter formed at RHIC using the full panoply of penetrating probes: jets, open and hidden charm, low-mass dileptons and photons 
. Many questions are still open, e.g. we have not yet seen unambiguous evidence for deconfinement or chiral symmetry 
restoration. These questions are high on the RHIC agenda and will be addressed in the next few runs. With the RHIC upgrades 
coming online and the LHC glimpsing over the horizon, I look forward to sustained progress and more exciting results over many 
years.

\section{Acknowledgements}
This work was supported by the Israel Science Foundation, the US-Israel Binational Science Foundation, the MINERVA Foundation 
and the Leon and Nella Benoziyo Center for High Energy Physics. I am grateful to David D'Enterria, Zeev Fraenkel, Peter Jacobs, 
Roy Lacey and Xin-Nian Wang for useful comments and discussions. Finally I wish to thank the QM05 organizers and in particular 
Peter Levai and Tamas Csorgo for such a wonderful and fruitful conference.

\end{document}